\documentclass{article}
\usepackage{arxiv}
\usepackage{dcolumn}
\usepackage{mathptmx} 
\usepackage{multirow}
\usepackage{}

\usepackage[utf8]{inputenc} % allow utf-8 input
\usepackage[T1]{fontenc}    % use 8-bit T1 fonts
\usepackage{hyperref}       % hyperlinks
\usepackage{url}            % simple URL typesetting
\usepackage{booktabs}       % professional-quality tables
\usepackage{amsfonts}       % blackboard math symbols
\usepackage{nicefrac}       % compact symbols for 1/2, etc.
\usepackage{microtype}      % microtypography
\usepackage{lipsum}
\usepackage{amssymb,amsmath,epsfig}
\newcommand{\beq}{\begin{equation}}
\newcommand{\eeq}{\end{equation}}
\newcommand{\bea}{\begin{eqnarray}}
\newcommand{\eea}{\end{eqnarray}}

\begin{document}
%\begin{flushright}
%AIP conference proceedings of the \\14th Asia-Pacific Physics Conference
%\end{flushright}
\title{Geodesics and optical properties of rotating black hole in Randall-Sundrum brane with a cosmological constant}

\author{{Saeed Ullah Khan\thanks{saeedkhan.u@gmail.com}} \, and \, {Jingli Ren\thanks{Corresponding author: renjl@zzu.edu.cn}}\vspace{0.2cm} \\\vspace{0.08cm}
Henan Academy of Big Data/School of Mathematics and Statistics, Zhengzhou University,
Zhengzhou 450001, China.}
\date{}
\maketitle
%%-------------------------------------------------%%
\begin{abstract}
The presence of tidal charge and a cosmological constant has considerable consequences on the spacetime geometry and its study is much important from the observational point of view. Henceforth, we investigate their effects on particle dynamics and the shadow cast by a Randall-Sundrum braneworld black hole with a cosmological constant. On studying the circular geodesics of timelike particles, we have acquired the expressions of energy, angular momentum and effective potential. We noted that the negative values of tidal charge and cosmological constant decreases the energy of particles. In addition, the negative value of cosmological constant leads us to the stable circular orbits, whereas its positive value destabilizes the circular orbits. Our exploration shows that the cosmological constant diminishes the radius of the black hole shadow. In response to the dragging effect, black hole rotation elongates its shadow toward the rotational axis. Besides, black hole spin and positive charge distort shadow and its distortion become maximum as far as the black hole rotates faster. We also discussed the energy emission rate by considering different cases and compared our result with the standard Kerr black hole.\\
\\
\end{abstract}
%\textbf{PACS:} 04.70.-s; 97.60.Lf; 04.70.Bw.
\tableofcontents
%\newpage
%\textbf{PACS:} 04.70.-s; 97.60.Lf; 04.70.Bw.
%%------------------------------------------------%%
\section{Introduction}
\label{intro}
%%----------------------------------------------------%%
In recent years, the investigation of higher-dimensional string and M-theories got a considerable interest of many researchers and are perhaps the aspiring approaches for the higher-
dimensional gravity theories \cite{Horava1,Horava2}. They influence the braneworld cosmological models \cite{Langlois,Peebles}, where the fields of gravity enter into 
the spatial extra dimensions. The existence of spatial extra dimensions can reform BH's properties and could have bigger 
size as compared to the Planck length scale ($l_p \sim 10^{-33}$cm) \cite{Arkani-Hamed}, while plays a dominant role in numerous gravity theories. Among these theories, the braneworld 
Randall–Sundrum (second) model is the simplest one, contains an extra spatial dimension, as well as a negative cosmological 
constant \cite{Randall}. As a result, such models could come up with an effective explanation of the hierarchy problems of electroweak and quantum gravity scales, while due to the huge 
extra-dimensional scales they can be of the same orders ($\sim$ TeV). Therefore, the future experiment could be used to test the braneworld models including the construction of hypothetical mini BHs of the TeV-energy scales \cite{Dimopoulos, Emparan}.

In the last few decades, particle dynamics around BHs have been an engaging problem in the field of astrophysics. The geodesic 
structure of a BH spacetime has been widely investigated by many researchers \cite{Stuchlik1,Tursunov1,Sharif1,Shahzadi}. 
As they are of great interest and could convey the eminent information and reveals the prolific structure of background 
geometry. Among various kinds of geodesics, the circular one is much engaging. They are very useful for the understanding and explaining of a BH's quasinormal modes (QNMs) \cite{Nollert}. By considering the non-zero cosmological constant the photon 
escape cone were successfully determined \cite{Hledik}. Pugliese et al. \cite{Pugliese1,Pugliese2} by examining the Reissner-Nordstr{\"o}m BHs conclude that the circular orbit will exist, even if the angular momentum is zero. Shaymatov et al. \cite{Shaymatov} explored particle dynamics around a 
spinning BH in the Randall-Sundrum brane with a cosmological constant. It is concluded that the Randall–Sundrum braneworld spacetimes can be stable against perturbations \cite{Toshmatov1}. Recently, Khan et al \cite{Saeed,Khan2} 
examined the motion of geodesics in the background of braneworld Kerr and a dyonic charged spacetimes, while a detailed study on the cosmic repulsion and external magnetic fields on accretion disks near spinning BHs could be found in \cite{Stuchlikk}. Deng \cite{Deng} explored the periodic orbits near a braneworld BH under the influence of tidal charge.

The shadow of a BH appears to be a dark region in the sky, which was for the first time introduced by Bardeen for the Kerr BH \cite{Bardeen}. His finding summarizes that over a background light source the radius of a BH shadow is $r_{shadow}=5.2M$. On investigating spinning BHs many researchers found that they cast shadows of almost the same size \cite{Takahashi,Hioki,Johannsen,Amarilla,Moffat}.
The theoretical exploration of a BH shadow could be described by the photons orbit and-or null geodesics. The shadows of spherically symmetric non-rotating BHs appeared to be circular, whereas the shadows of rotating BHs appeared in deformed shape \cite{Perlick}. In recent years, the astrophysical advances attract many researchers to explore BH shadows \cite{Vries,Abdu1,Haroon,Konoplya}. It is accepted that in the very near future BH shadows can directly be investigated \cite{Cash,Doeleman,Falcke,Bambi}. As a result, the exploration of BH's shadow would be a rewarding technique to understand the astrophysical BHs and to compare the general theory of relativity with modern theories \cite{Eiroa}. Khan and Ren \cite{Khan3} deduced that  shadow radius increases with the increase of quintessence parameter, while the charge and spin parameters result in distortion of the BH shadow. The shadow of braneworld BH is investigated under the influence of the brane parameter \cite{Neves}. Recently with the help of Event Horizon Telescope (EHT), scientists have successfully obtained the supermassive BH's image at the middle of the M87 galaxy \cite{ETH,Akiyama1,Akiyama2}. On considering the Randall-Sundrum braneworld BH, it is shown that the observation of M87{*}’s dark shadow sets the limit $\ell \lesssim 170$AU (where $\ell$ denote the AdS$_{5}$ curvature radius) \cite{Vagnozzi}. Such solid evidence about the existence of supermassive BHs exposes new ways to the exploration of BHs.

This article explores the circular geodesic motion around a rotating braneworld BH with a cosmological constant and its shadow. In section 2, we will briefly review the said BH and the structure of its horizons. By considering the timelike geodesics we will discuss energy, angular momentum and the effective potential of particles in section 3. Section 4 will provide a thorough discussion of the shadow and energy emission of the said BH. In the last section, we will thorough discuss and summarize our obtained results.
%%----------------------------------------------------------%%
\section{Rotating braneworld black hole with a cosmological constant}
%%----------------------------------------------------------%%
This section aimed to investigate the rotating braneworld black hole with a cosmological constant, obtained by Neves and Molina \cite{Neves}. The spacetime metric of the said BH in Boyer-Lindquist coordinates $(t, r,\phi,\theta)$, simplifies to
\begin{eqnarray}\label{e1}\nonumber
ds^2&=&-\frac{1}{\rho^2}(\Delta_r-\Delta_{\theta} a^2 \sin^2\theta)dt^2-\frac{2a}{\Xi\rho^2}[(a^2+r^2)\Delta_\theta-\Delta_r]\sin^2\theta d\phi dt +\\
&& \frac{\rho^2}{\Delta_\theta}d\theta^2+\frac{\rho^2}{\Delta_r}dr^2+\frac{1}{\Xi^2 \rho^2}[(a^2+r^2)^2\Delta_\theta-\Delta_r a^2\sin^2\theta]\sin^2\theta d\phi^2 ,
\end{eqnarray}
with
\begin{eqnarray}\nonumber
\Delta_r &= &(a^2+r^2)\left(1-\frac{r^2}{3}\Lambda_{4D}\right)-2Mr+q, \quad \Xi=1+\frac{a^2}{3}\Lambda_{4D}, \\\nonumber
\rho^2 &=&r^2+a^2 \cos^{2}\theta, \quad \Delta_\theta = 1+ \frac{a^2}{3}\Lambda_{4D} \cos^2\theta.
\end{eqnarray}
Here $M$, $q$ and $a$, respectively denote mass, tidal charge and spin parameter of the BH, while $\Lambda_{4D}$ is the four-dimensional brane cosmological constant. The above model \eqref{e1}, simplifies to the KN spacetime by letting $\Lambda_{4D}=0$ and replacing $q$ by $Q^2$ (representing electric charge of the KN BH). Model \eqref{e1}, reduces to the RN BH if $\Lambda_{4D}=a=0$; to the Kerr BH \cite{Kerr} if $\Lambda_{4D}=q=0$; and finally to the Schwarzschild BH \cite{Schwarz} by letting $\Lambda_{4D}=a=q=0$.
The horizons of metric \eqref{e1}, can be obtained from the solution of $\Delta_r=0$, as
\beq\label{EH}
(a^2+r^2)\left(1-\frac{r^2}{3}\Lambda_{4D}\right)-2Mr+q=0.
\eeq
%%-----------------------------------------------------%%
\begin{figure*}
\begin{minipage}[b]{0.58\textwidth} \hspace{-0.4cm}
        \includegraphics[width=0.8\textwidth]{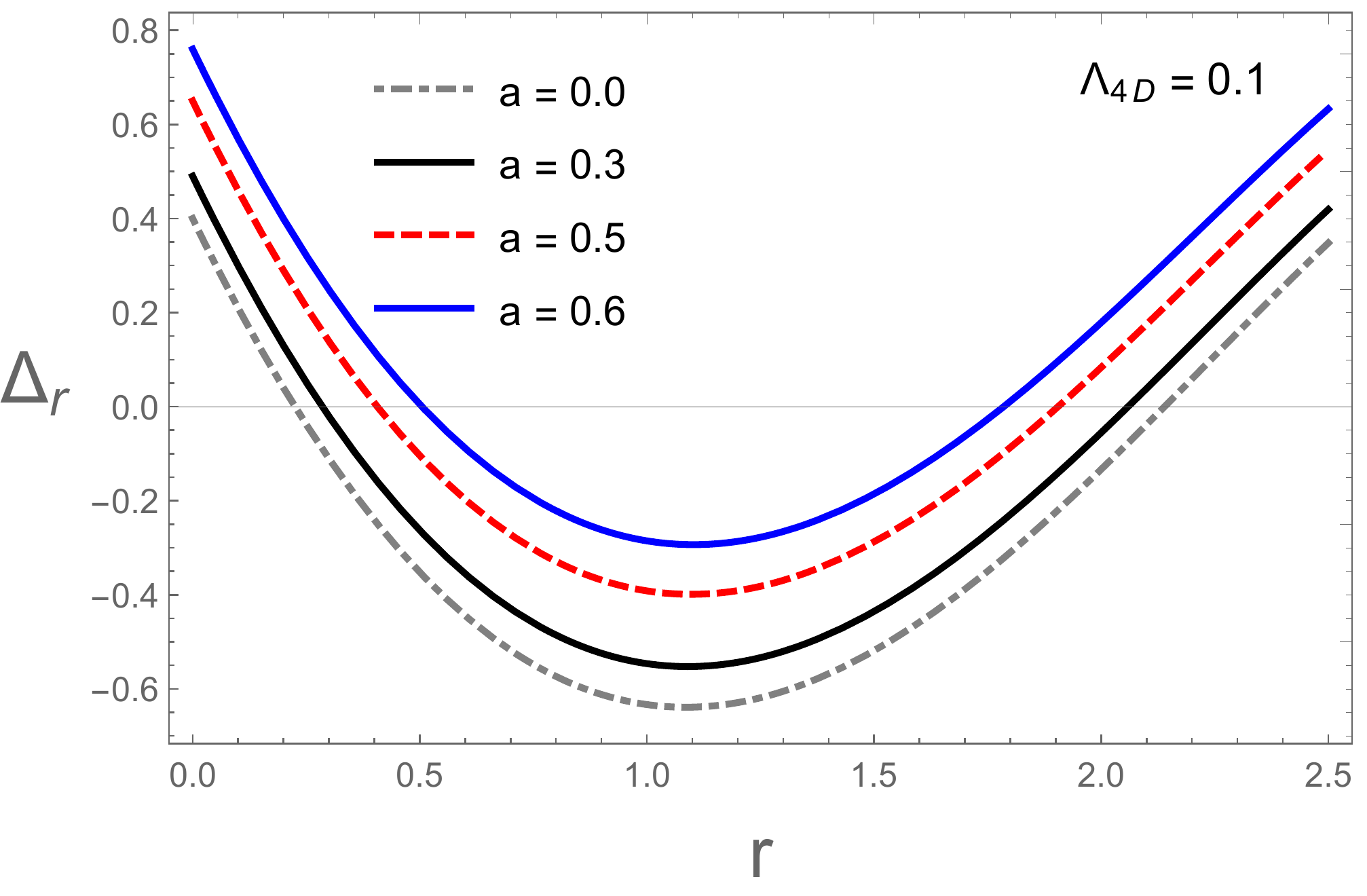}
    \end{minipage}
    \vspace{0.3cm}
        \begin{minipage}[b]{0.58\textwidth} \hspace{-1.2cm}
       \includegraphics[width=.8\textwidth]{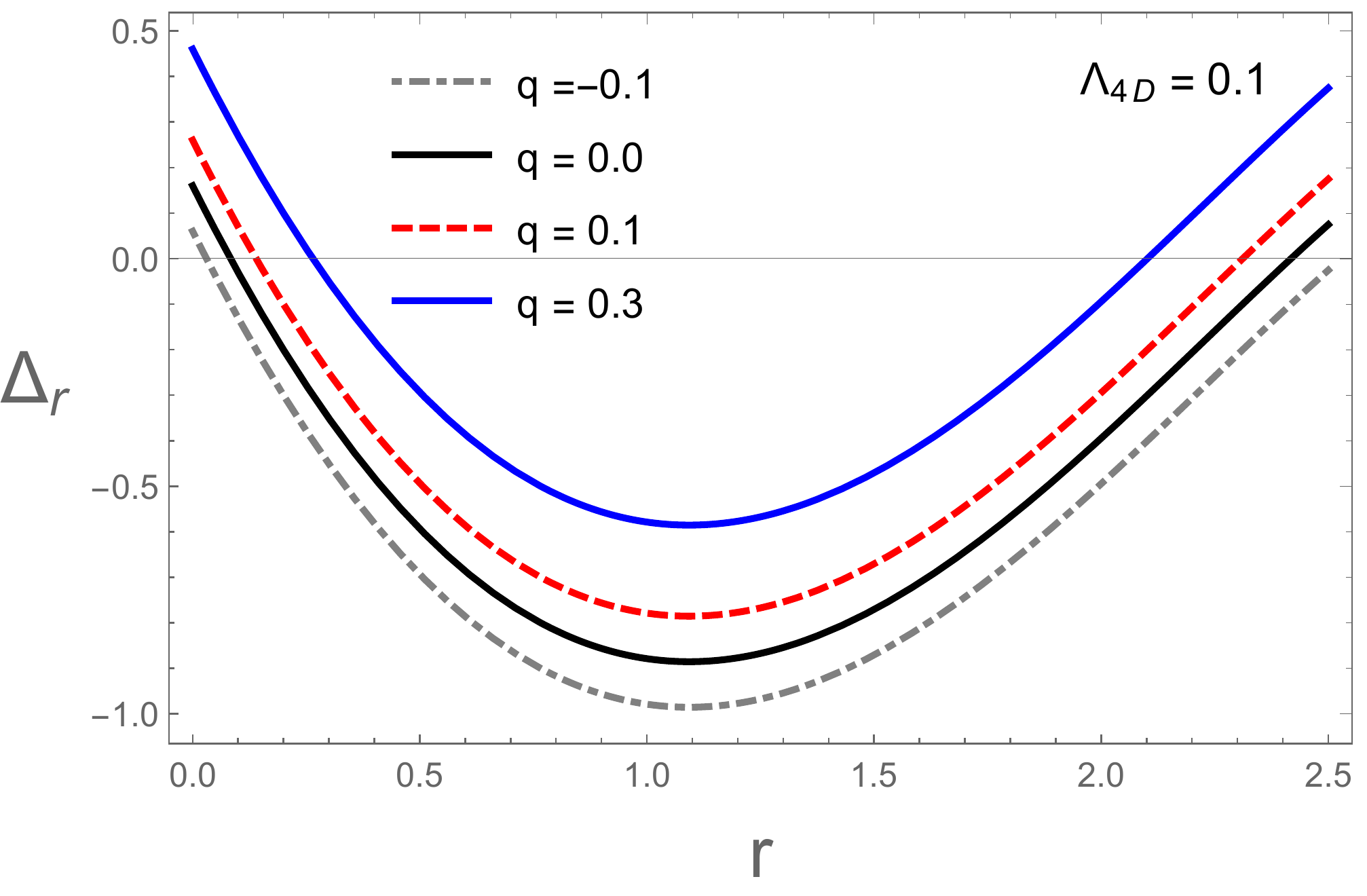}
    \end{minipage}
\begin{minipage}[b]{0.58\textwidth} \hspace{-0.4cm}
        \includegraphics[width=0.8\textwidth]{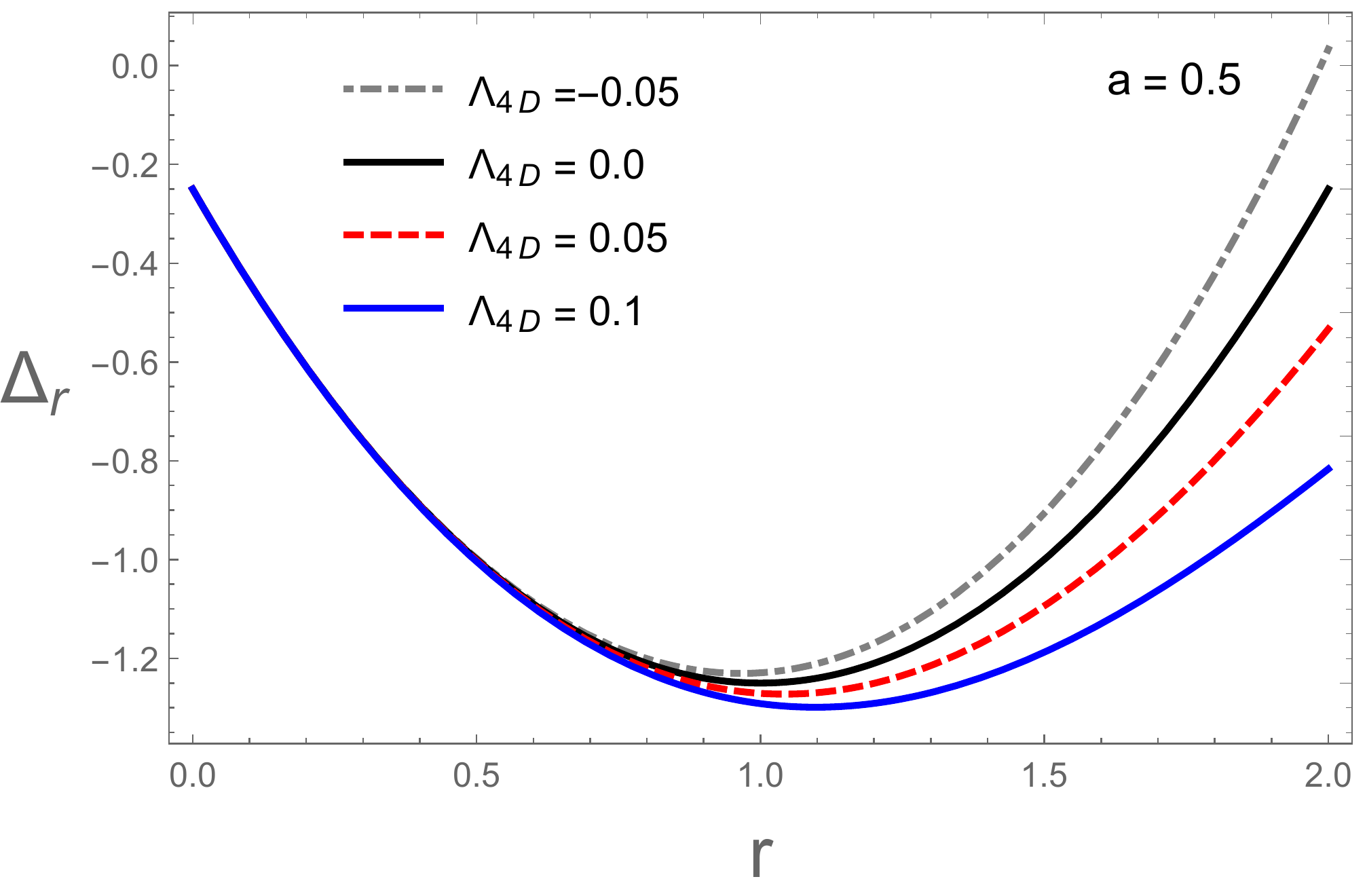}
    \end{minipage}
        \begin{minipage}[b]{0.58\textwidth} \hspace{-1.2cm}
       \includegraphics[width=.8\textwidth]{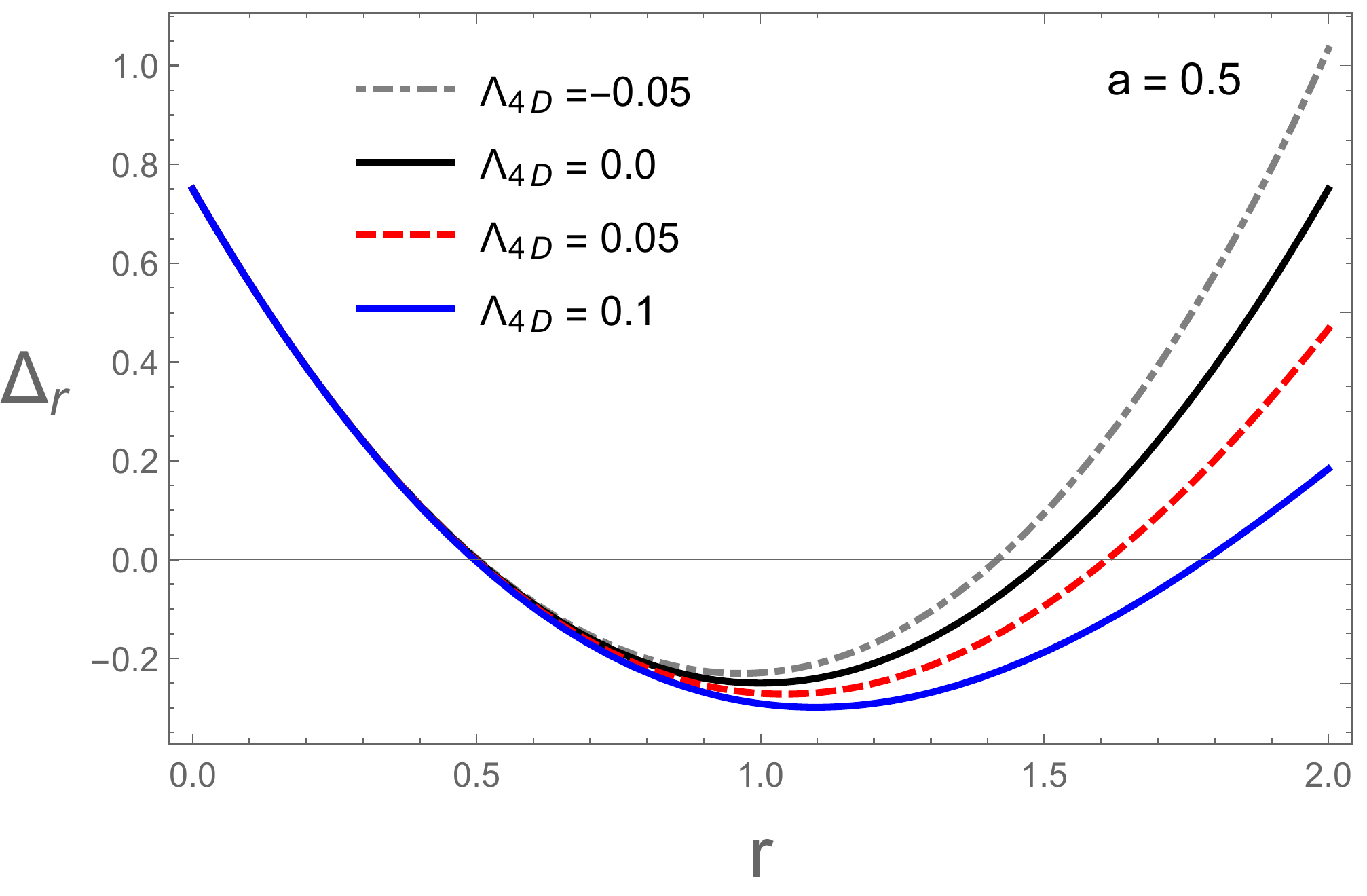}
    \end{minipage}
\caption{The radial profile of BH horizons in the upper row at $q=0.4$ (left) and $a=0.4$ (right), while the lower row is plotted for $q=-0.5$ (left) and $q=0.5$ (right).}\label{horizons}
\end{figure*}
%%-----------------------------------------------------%%
On the other hand, the ergosphere can be obtained by letting $g_{tt} = 0$, as
\beq\label{SLS}
a^2\sin^2{\theta \Delta_{\theta}}-\Delta_r=0,
\eeq
which represents the boundary of static limit surface (SLS), while in the absence of cosmological constant, the SLS takes the form
\beq\label{SLS1}
r_{SLS\pm}=M\pm \sqrt{M^2-a^2\cos^2{\theta}-q}.
\eeq
The above Eq. \eqref{SLS1}, shows that besides other parameters the SLS also depends on $\theta$. The numerical exploration of the event horizon and SLS vs the radial distance $r$ is described in Figs. \ref{horizons} and \ref{SLS}. We noted that in case of fast-rotating BH the area of event horizons get reduced. The negative value of tidal charge increases the event horizons as well as SLS, while it's positive value results in a decrease of the SLS and event horizon. On the other hand, $\Lambda_{4D}<0$ diminishing the event horizon, whereas $\Lambda_{4D}>0$ contributes to the area of the event horizon. Moreover, the positive value of cosmological constant increase the SLS and $\Lambda_{4D}<0$ results in a decrease of the SLS.
%%-----------------------------------------------------%%
\begin{figure*}
\begin{minipage}[b]{0.58\textwidth} \hspace{-0.4cm}
        \includegraphics[width=0.8\textwidth]{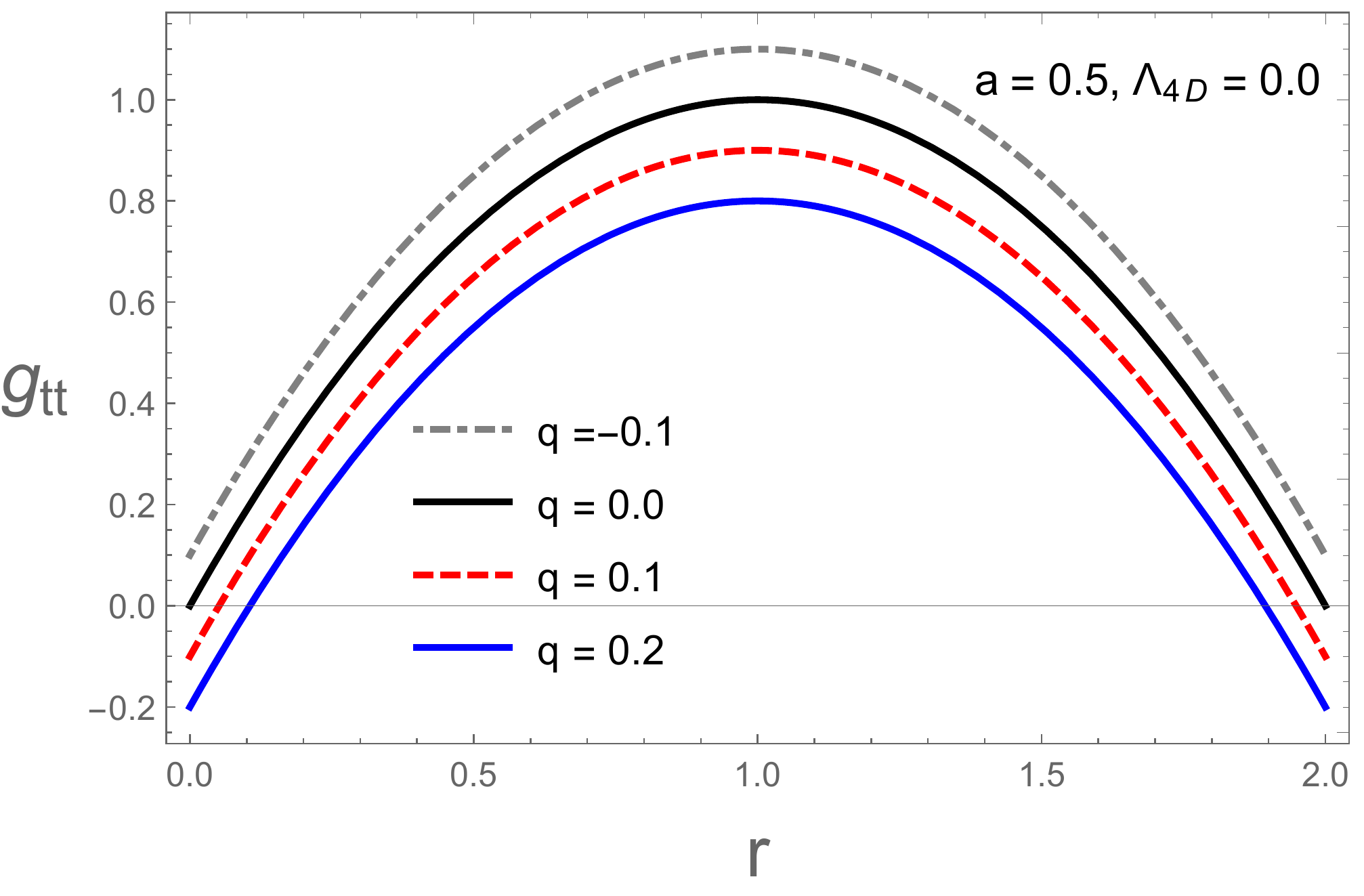}
    \end{minipage}
    \vspace{0.3cm}
        \begin{minipage}[b]{0.58\textwidth} \hspace{-1.2cm}
       \includegraphics[width=.8\textwidth]{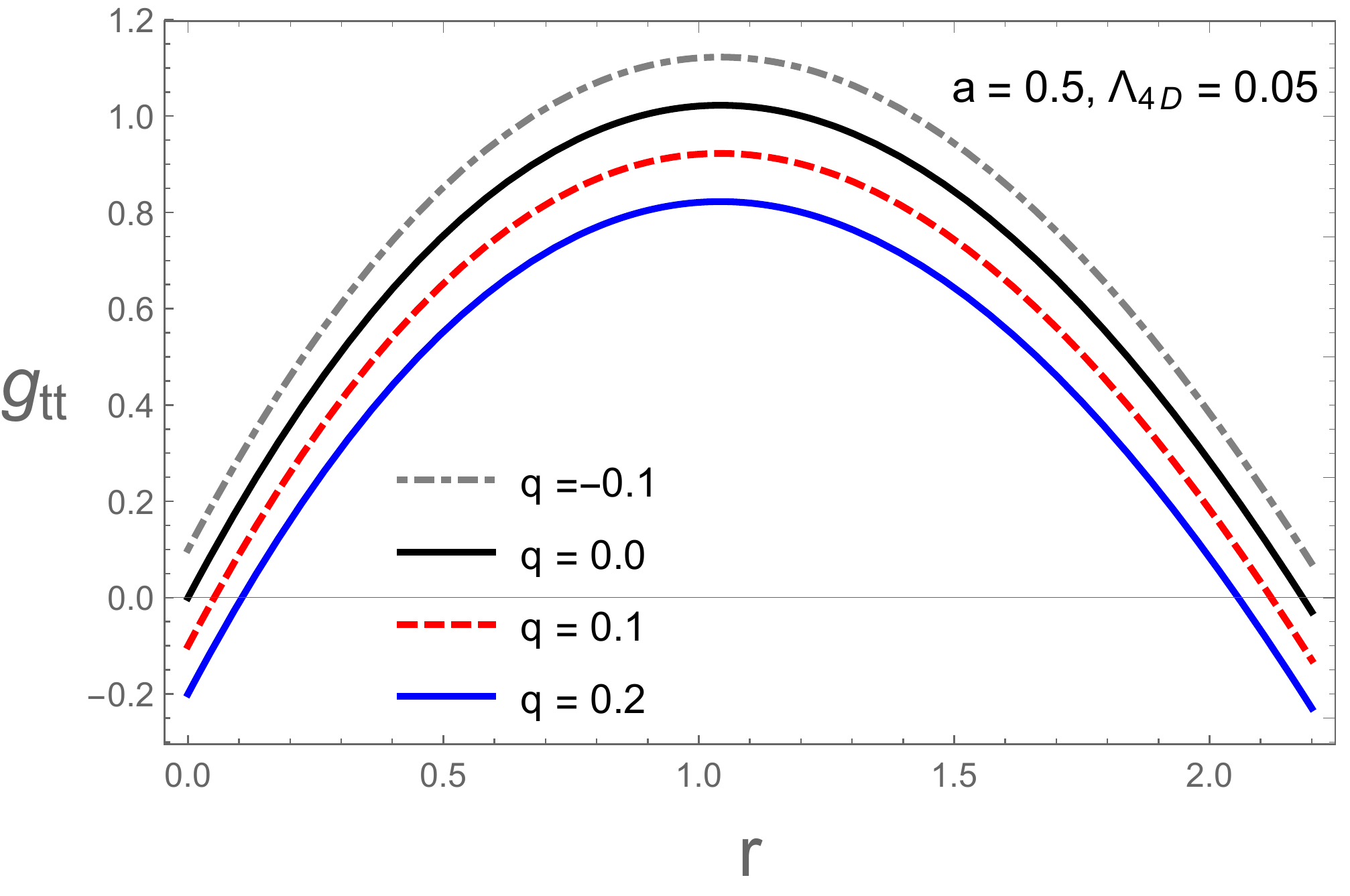}
    \end{minipage}
\begin{minipage}[b]{0.58\textwidth} \hspace{-0.4cm}
        \includegraphics[width=0.8\textwidth]{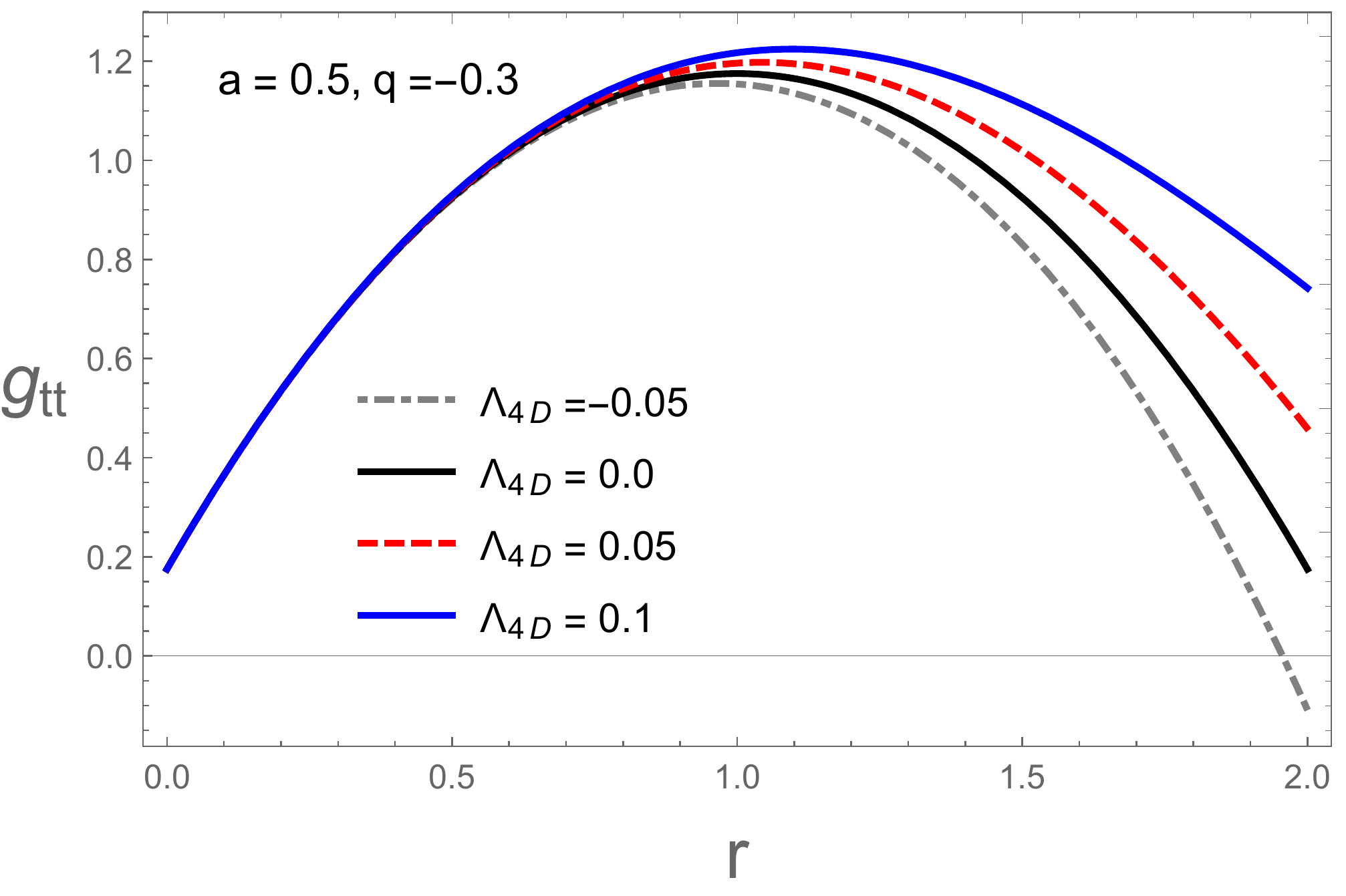}
    \end{minipage}
        \begin{minipage}[b]{0.58\textwidth} \hspace{-1.2cm}
       \includegraphics[width=.8\textwidth]{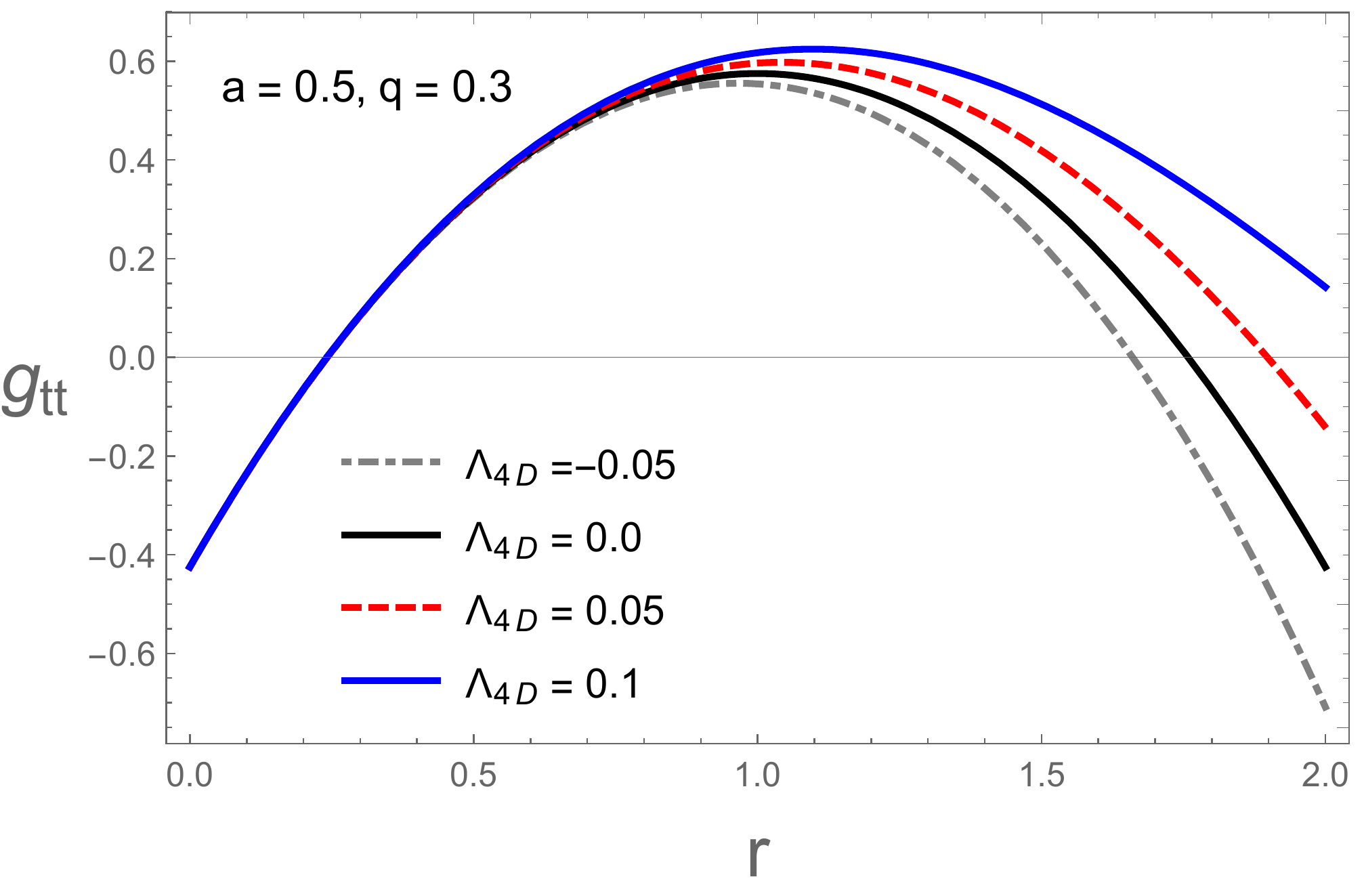}
    \end{minipage}
\caption{The radial profile of SLS in the upper row at $\theta=\pi/2$, while in the lower row at $\theta=\pi/4$.}\label{SLS}
\end{figure*}
%%-------------------------------------------------%%
\section{Circular geodesics}
%%-------------------------------------------------%%
Test particle motion around a Randall-Sundrum braneworld BH with a cosmological constant could be outlined using the Lagrangian equation by
\begin{equation}\label{G1}
\mathcal{L} = \frac{1}{2}g_{\mu\nu}\dot{x}^{\mu}\dot{x}^{\nu},
\end{equation}
in the above equation, dot represents derivative with respect to the proper time ($\tau$) where $\dot{x}^{\mu}={dx^{\mu}}/{d\tau}$, while $x^\mu$ stands for the corresponding four velocities of particle. Using BH symmetry on the equatorial plane, we can define the following conserved quantities
\begin{eqnarray}\label{G2}
-\mathcal{E}&=&-u_{t}=g_{tt}\dot{t}+g_{t\phi} \dot{\phi},\\ \label{G3}
L_z&=&u_{\phi}=g_{t\phi}\dot{t}+g_{\phi\phi}\dot{\phi}.
\end{eqnarray}
Here $\mathcal{E}$ and $L$, respectively represent the total energy and angular momentum of particles. On solving the above two equation for timelike geodesics ($u^{\mu}u_{\mu}=-1$), we can obtain
\begin{eqnarray}\label{G4}
\frac{dt}{d\tau}&=&-\frac{1}{r^2}\left[ a(a\mathcal{E}- \Xi L)-\frac{(a^2+r^2)}{\Delta_r}\left((a^2+r^2)\mathcal{E}-\Xi aL \right)\right],\\\label{G5}
\frac{d\phi}{d\tau}&=&-\frac{1}{r^2} \left[ (a\mathcal{E}-\Xi L \right)-\frac{a}{\Delta_r}\left((a^2+r^2)\mathcal{E}-\Xi aL) \right].
\end{eqnarray}
Hence, the equations of motion for a particle take the form \cite{Shaymatov}
\begin{eqnarray}\label{G6}
\dot{r}^2=\left(\frac{dr}{d \tau}\right)^2=\frac{1}{r^4}\left[\left((a^2+r^2)\mathcal{E}-\Xi aL\right)^2-\Delta_r \left(r^2+(\Xi L-a\mathcal{E})^2 \right) \right].
\end{eqnarray}
By considering the special case of $a \mathcal{E}=\Xi L$, Eq. \eqref{G6} simplifies to
\beq\label{G7}
\dot{r}^2=\mathcal{E}^2-\frac{\Delta_r}{r^2}.
\eeq
On integrating with respect to $r$, the above equation reduces to
\beq\label{G8}
\tau=\int\left(\mathcal{E}^2-\frac{\Delta_r}{r^2} \right)^{-\frac{1}{2}} dr.
\eeq
Next, we are interested in a general case where $\Xi L \neq a\mathcal{E}$, therefore, by substituting $x = \Xi L-a\mathcal{E}$ in Eq. \eqref{G6}, we get
\beq\label{G9}
\mathcal{F}(r)=r^4 \dot{r}^2=(a^2-\Delta_r)x^2-2ar^2\mathcal{E}x+r^4\mathcal{E}^2-r^2\Delta_r.
\eeq
On taking derivative with respect to r, Eq. \eqref{G9} takes the form
\beq\label{G10}
\mathcal{F^\prime}(r)=-\Delta_r^\prime x^2-4ar\mathcal{E}x+4r^3\mathcal{E}^2-2r\Delta_r-r^2\Delta_r^\prime.
\eeq
On substituting both $\mathcal{F}(r)$ and $\mathcal{F^\prime}(r)$ equal to zero and then by equating them, we can obtain
\beq\label{G11}
r^4 \mathcal{E}^2-\frac{r^3\Delta_r^{\prime}}{2}+ \left(\Delta_r- a^2-\frac{r \Delta_r^{\prime}}{2}\right) x^2=0.
\eeq
\beq\label{G12}
\mathcal{E}= \frac{1}{ar^2x}\left(\left(\frac{r \Delta_r^{\prime}}{4}+a^2-\Delta_r \right) x^2-\frac{r^2\Delta_r}{2} +\frac{r^3\Delta_r^{\prime}}{4}\right).
\eeq
The elimination of $\mathcal{E}$, from the above two equations, leads us to the following quadratic equation in $x^2$
\beq\label{G13}
Ax^4+Bx^2+C=0.
\eeq
Here
\begin{eqnarray}\label{G14}\nonumber
A&=&4\left(\frac{r \Delta_r^{\prime}}{4}+a^2-\Delta_r \right)^2+a^2(4\Delta_r-2r\Delta_r^{\prime}-4a^2),\\ \nonumber
B&=&(4a^2+r\Delta_r^{\prime}-4\Delta_r)\left(\frac{r^3\Delta_r^{\prime}}{2}-r^2\Delta_r \right)-2a^2r^3\Delta_r^{\prime},\\ \nonumber
C&=&\left(\frac{r^3\Delta_r^{\prime}}{2}-r^2\Delta_r \right)^2.
\end{eqnarray}
Solving Eq. \eqref{G13} for $x^2$, result in
\beq\label{G14}
x^2=\frac{-B\pm\sqrt{D}}{2A},
\eeq
with
\beq\label{G14}
D=16a^2r^4 \Delta_r^2 \left(a^2+\frac{r \Delta_r^{\prime}}{2}-\Delta_r \right) .
\eeq
Consequently, the solution of Eq. \eqref{G13}, can be acquired as
\beq\label{G15}
x=\frac{-r}{\sqrt{\mathcal{Z_{\mp}}}}\left(a\pm \sqrt{\frac{r\Delta_r^\prime}{2}+a^2-\Delta_r}\right),
\eeq
in which
\beq\label{G16}
\mathcal{Z_{\mp}}= 2(\Delta_r-\frac{r\Delta_r^\prime}{4}-a^2)\mp a\sqrt{2r\Delta_r^\prime-4(\Delta_r-a^2)} .
\eeq
The upper and lower signs respectively correspond to the counter-rotating and co-rotating orbits. Substituting the value of $x$ into Eq. \eqref{G11}, the energy of the circular orbit, can be obtained as
\beq\label{G17}
\mathcal{E}=\frac{1}{r\sqrt{\mathcal{Z\mp}}}\left(\Delta_r-a^2\mp a\sqrt{\frac{r\Delta_r^\prime}{2}+a^2-\Delta_r }\right).
\eeq
By utilizing the above equation, the corresponding angular momentum simplifies to
\beq\label{G18}
L=\frac{1}{r\sqrt{\mathcal{Z\mp}}}\left(a(\Delta_r-r^2-a^2)\mp (r^2+a^2)\sqrt{\frac{r\Delta_r^\prime}{2}+a^2-\Delta_r }\right).
\eeq
The above Eqs. \eqref{G17} and \eqref{G18}, respectively describe the energy and angular momentum of a particle's circular orbit, while the test particle's angular velocity can be defined as
\beq\label{G18a}
\Omega_{\mp}=\frac{d\phi}{dt}=\frac{\mp \sqrt{\frac{r\Delta_r^\prime}{2}+a^2-\Delta_r }}{r^2 \mp a\sqrt{\frac{r\Delta_r^\prime}{2}+a^2-\Delta_r }}.
\eeq
In Eq. \eqref{G18a}, the upper and lower signs respectively correspond to the counter-rotating and co-rotating orbits. Using Eqs. \eqref{G17} and \eqref{G18}, one can find two reality conditions on the existence of circular orbits. The first reality condition could be expressed as
\beq\label{G19}
\mathcal{Y} \leq \mathcal{Y}_s \equiv \frac{r^{-4}}{2}\left(-2q+2r \right).
\eeq
In which $\mathcal{Y}={M^2\Lambda_{4D}}/{3}$ and for simplicity, we take $M=1$. Eq. \eqref{G19}, initiate the concept of "{\it static radius}" which can be acquired from
\beq\label{G20}
-2q+2r=2r^4 \mathcal{Y},
\eeq 
while the second condition on the existence of circular orbits takes the form
\beq\label{G21}
2(\Delta_r-\frac{r\Delta_r^\prime}{4}-a^2)\mp a\sqrt{2r\Delta_r^\prime-4(\Delta_r-a^2)} \geq 0.
\eeq
%%-----------------------------------------------------%%
\begin{figure*}
\begin{minipage}[b]{0.58\textwidth} \hspace{-0.2cm}
\includegraphics[width=0.8\textwidth]{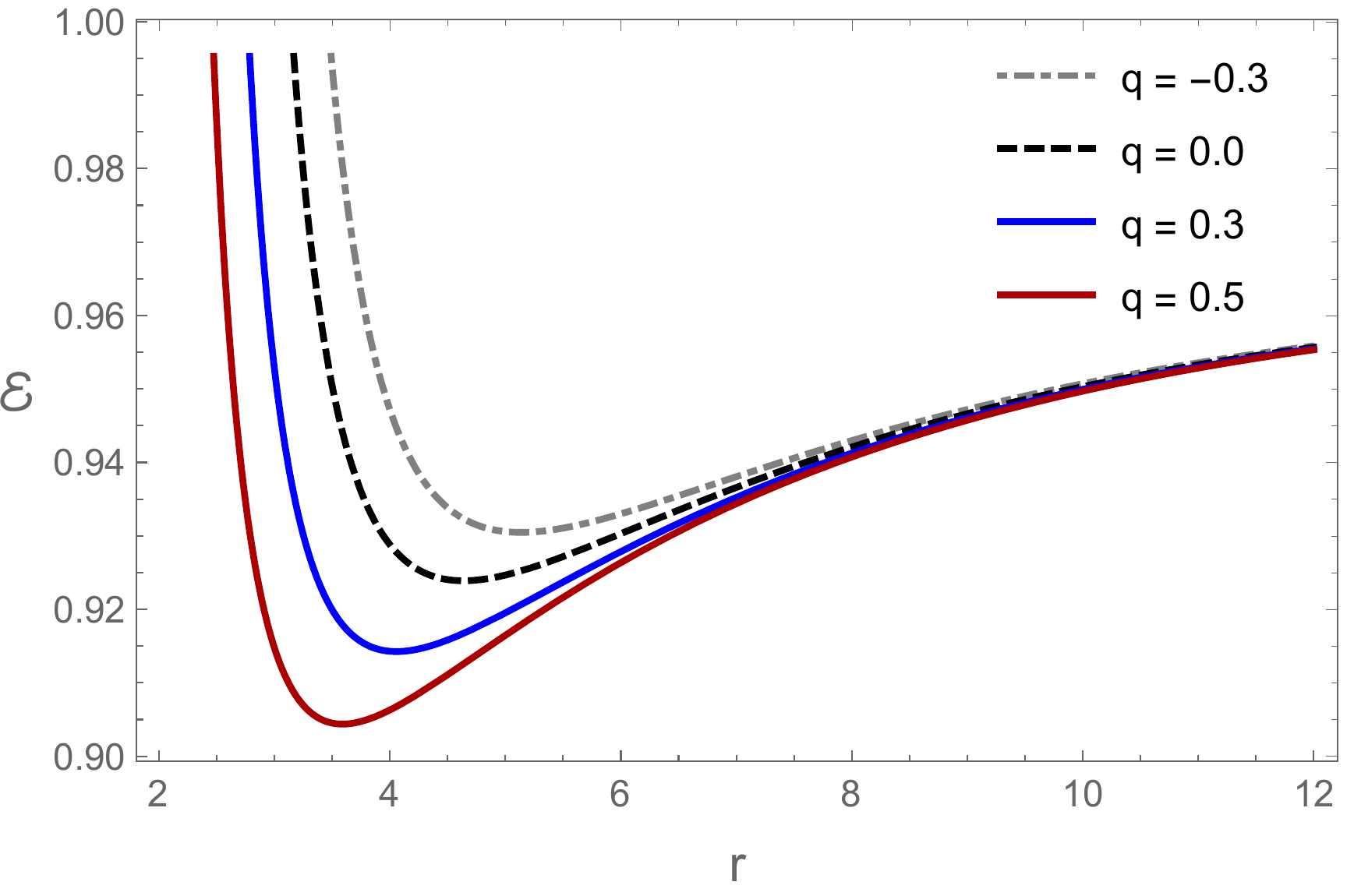}
\end{minipage}\vspace{0.3cm}
\begin{minipage}[b]{0.58\textwidth} \hspace{-1.1cm}
\includegraphics[width=0.8\textwidth]{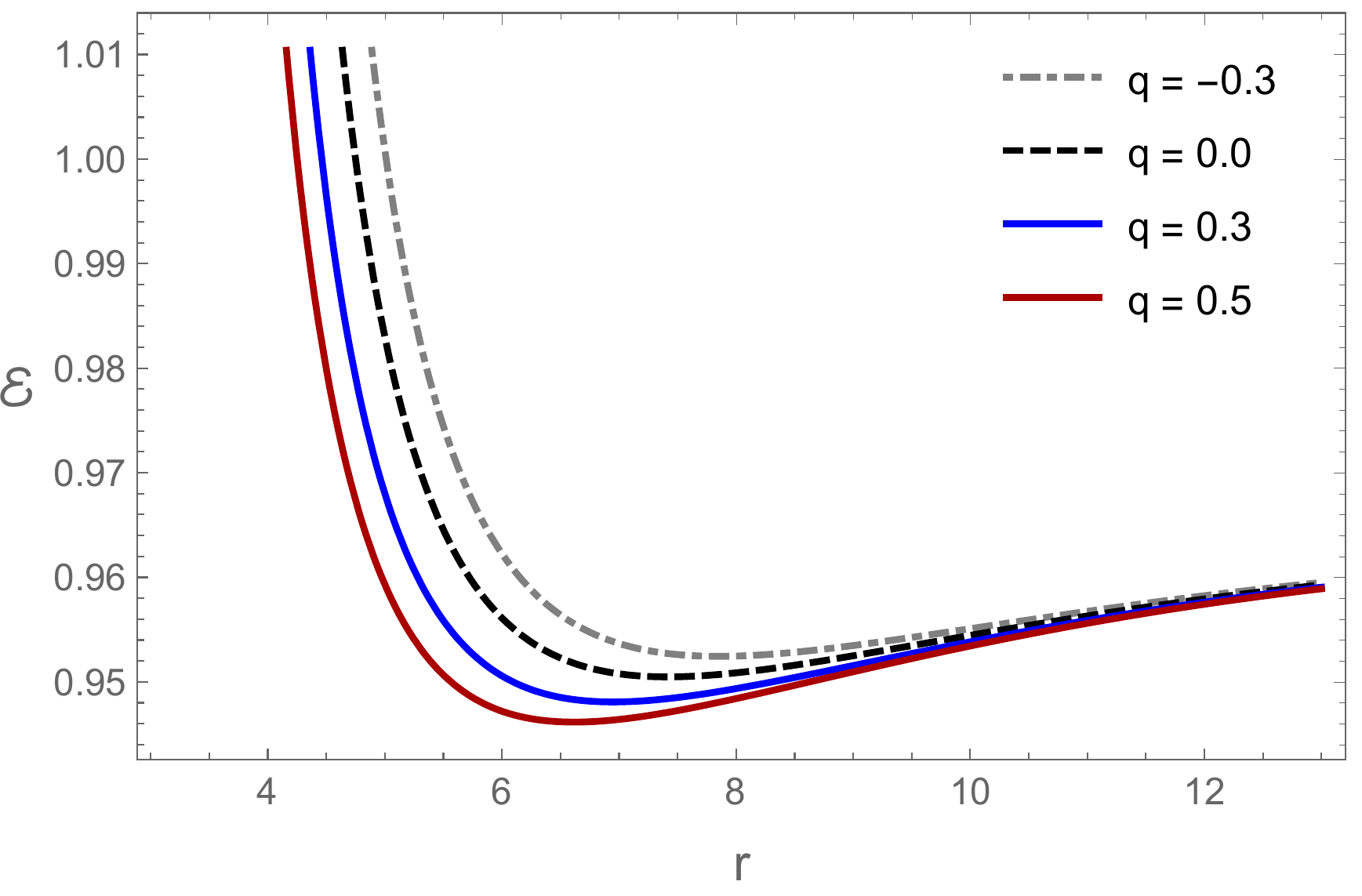}
\end{minipage}
      %%----------%%
\begin{minipage}[b]{0.58\textwidth} \hspace{-0.2cm}
\includegraphics[width=0.8\textwidth]{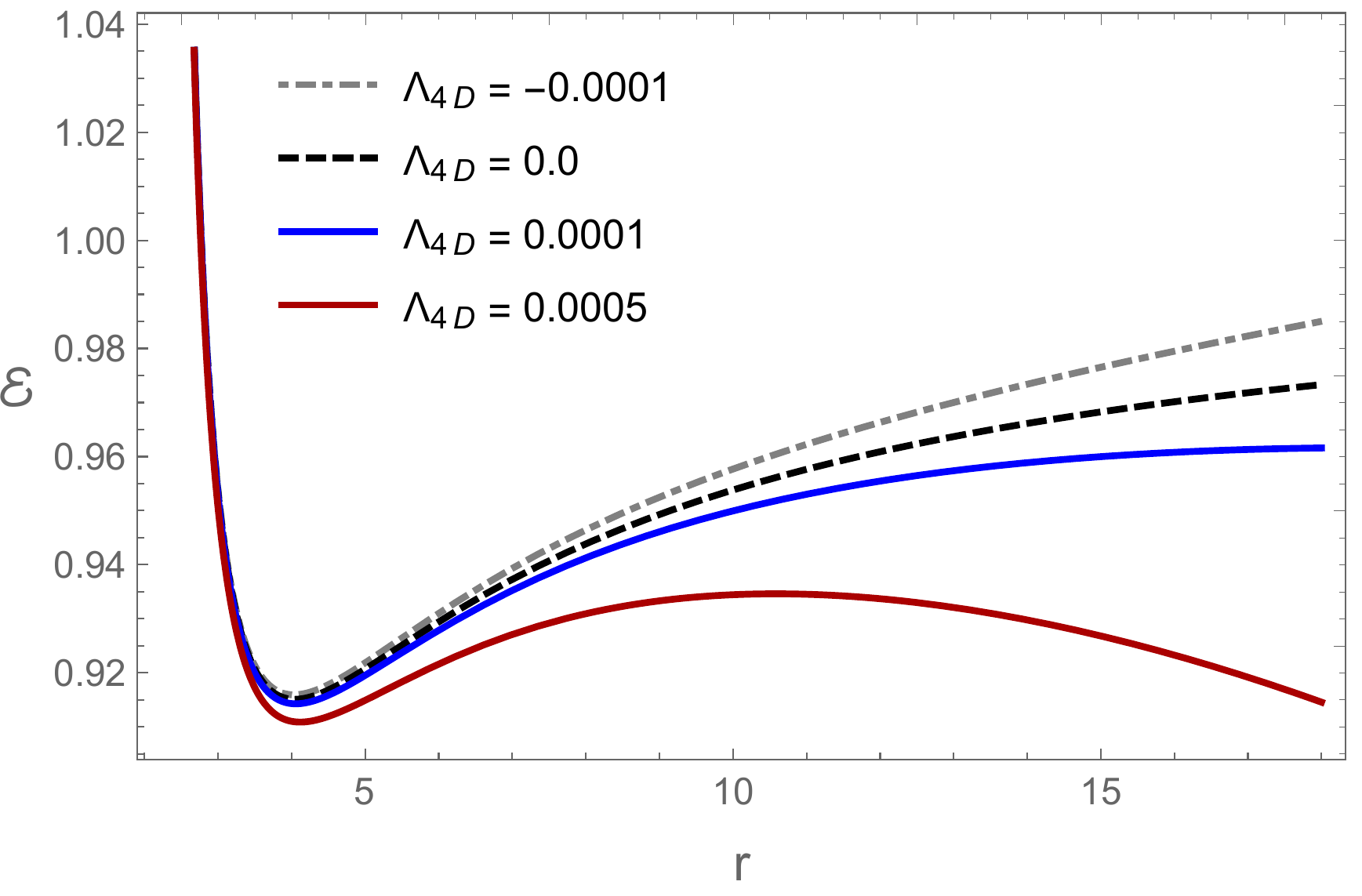}
\end{minipage}
\begin{minipage}[b]{0.58\textwidth}\hspace{-1.1cm}
\includegraphics[width=0.8\textwidth]{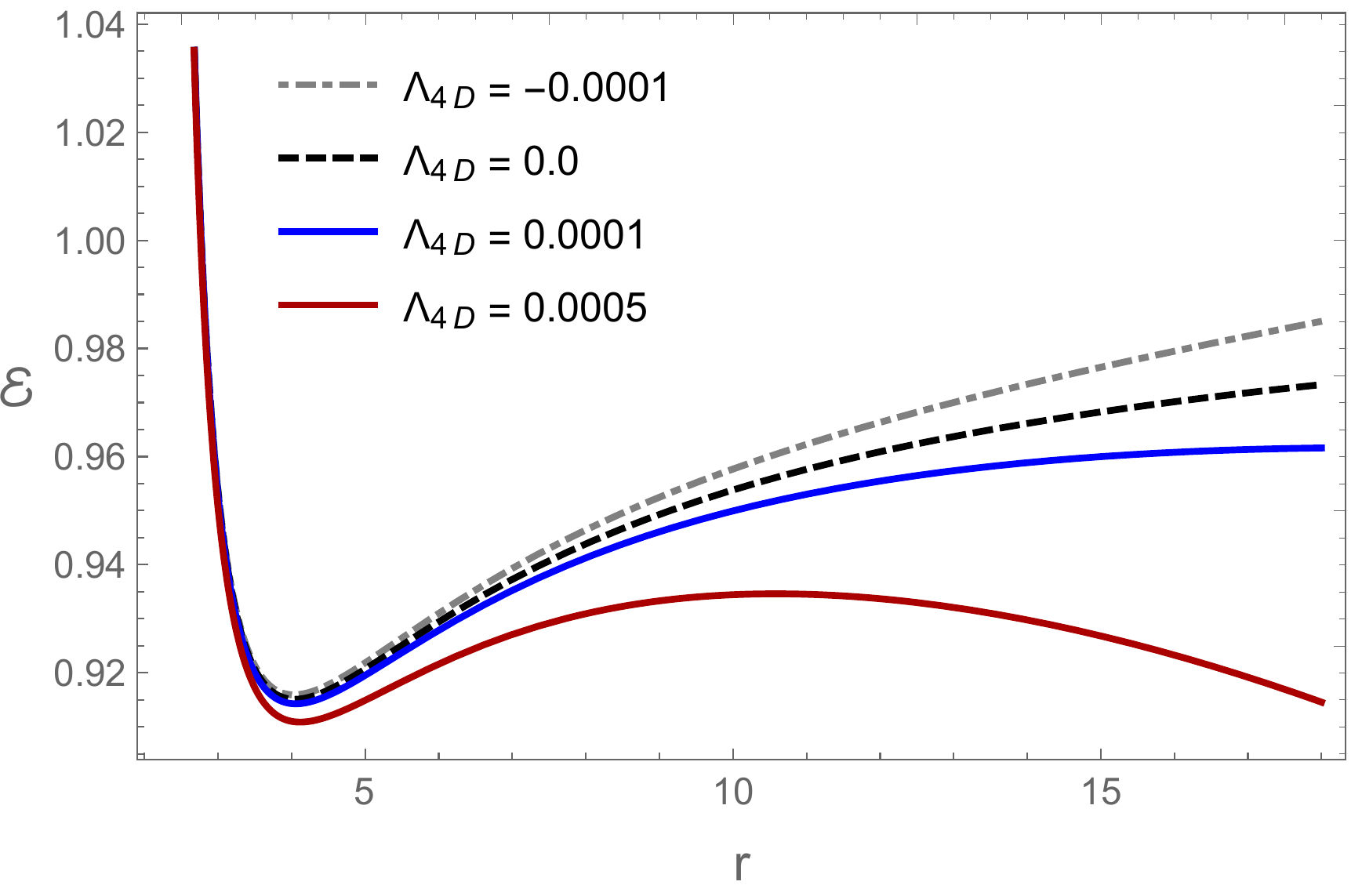}
\end{minipage}
\caption{Radial profile of the circular orbit's energy $\mathcal{E}$, left for prograde and right for retrogade particles at $a=0.4$ and $\Lambda_{4D}=0.0001$ (top row), while at $a=0.4$ and $q=0.3$ (bottom row).}\label{Energy}
\end{figure*}
%%----------------------------------------------------%%
%%-----------------------------------------------------%%
\begin{figure*}
\begin{minipage}[b]{0.58\textwidth} \hspace{-0.2cm}
\includegraphics[width=0.8\textwidth]{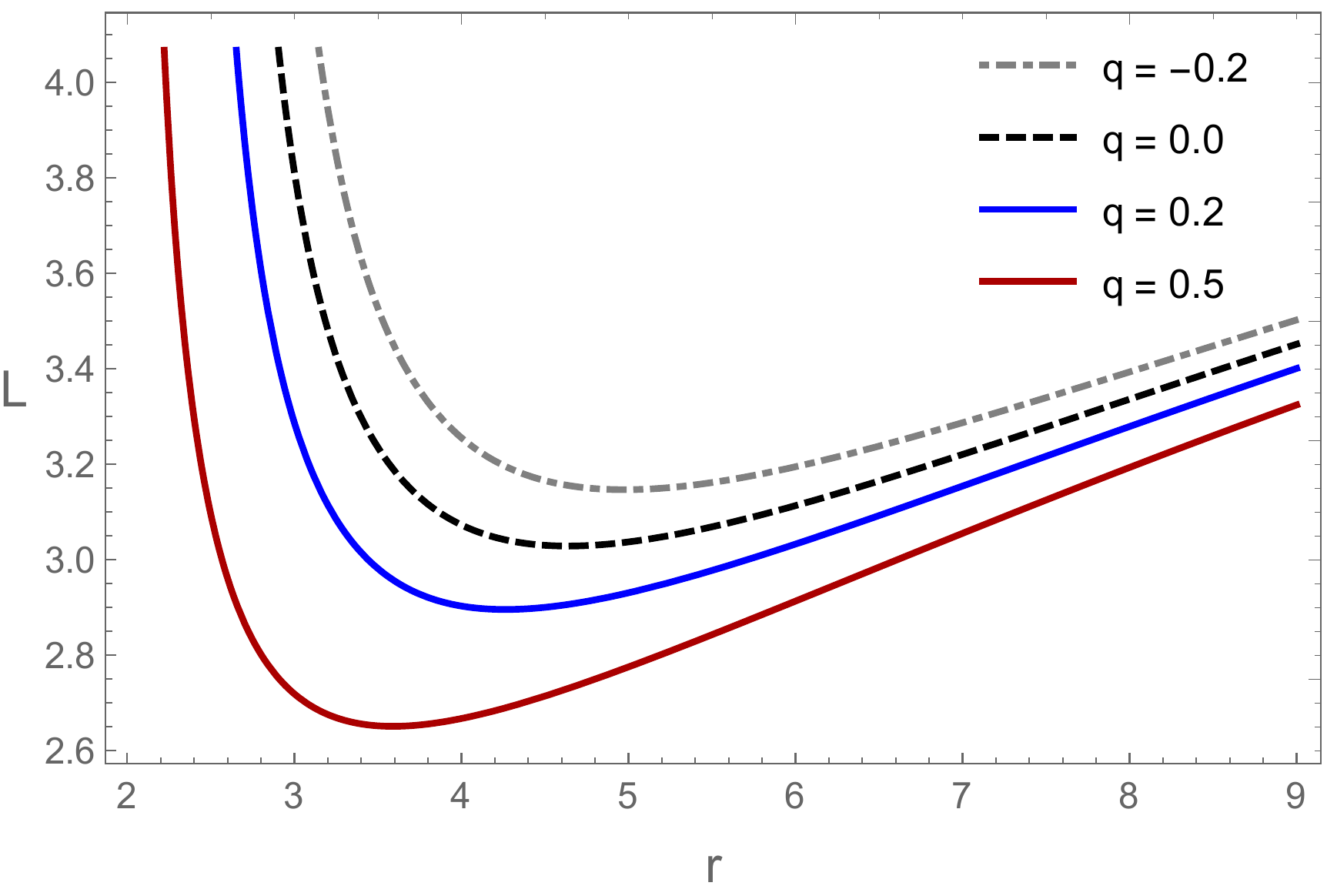}
\end{minipage}\vspace{0.3cm}
\begin{minipage}[b]{0.58\textwidth} \hspace{-1.1cm}
\includegraphics[width=0.8\textwidth]{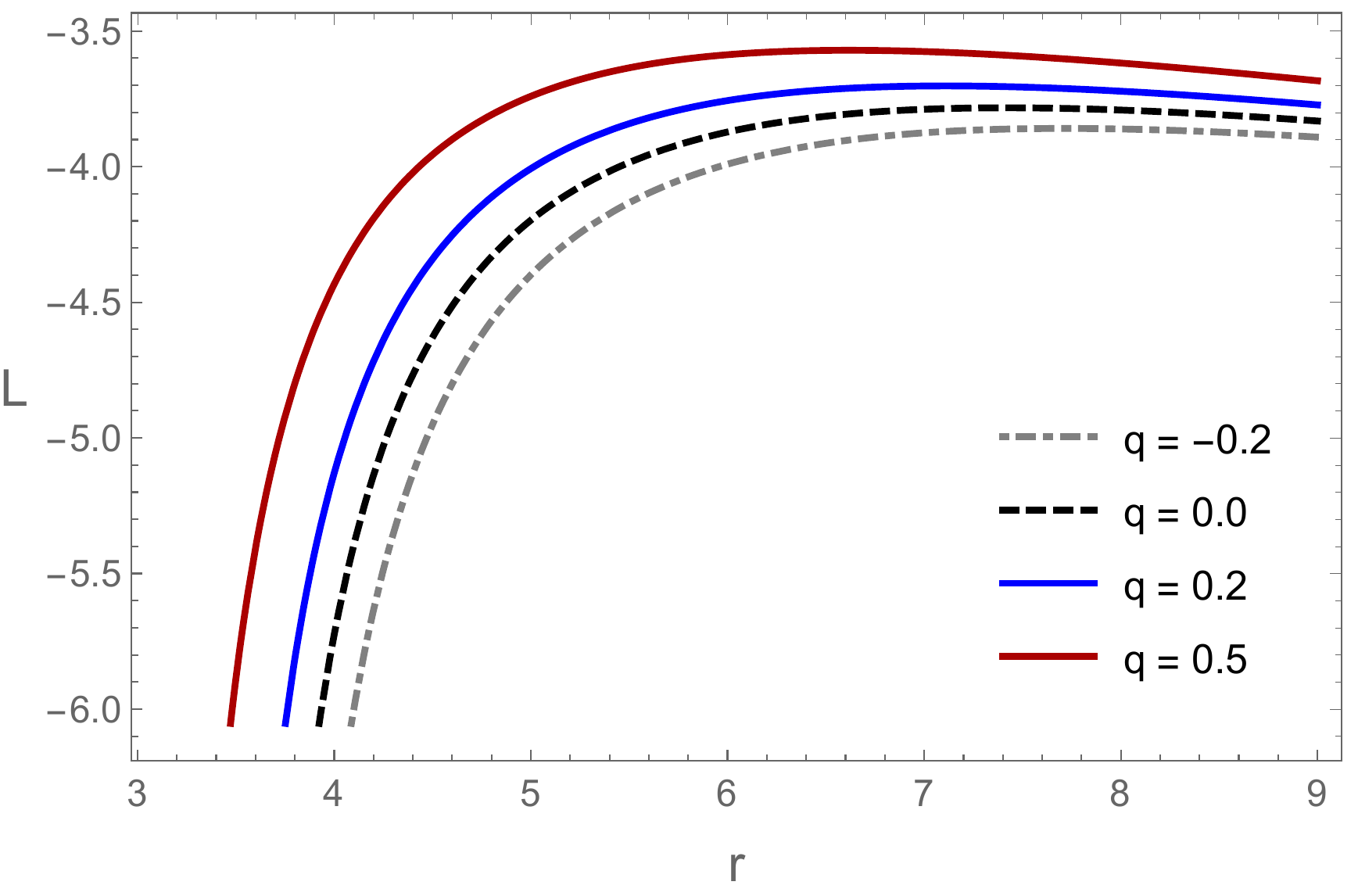}
\end{minipage}
      %%----------%%
\begin{minipage}[b]{0.58\textwidth} \hspace{-0.2cm}
\includegraphics[width=0.8\textwidth]{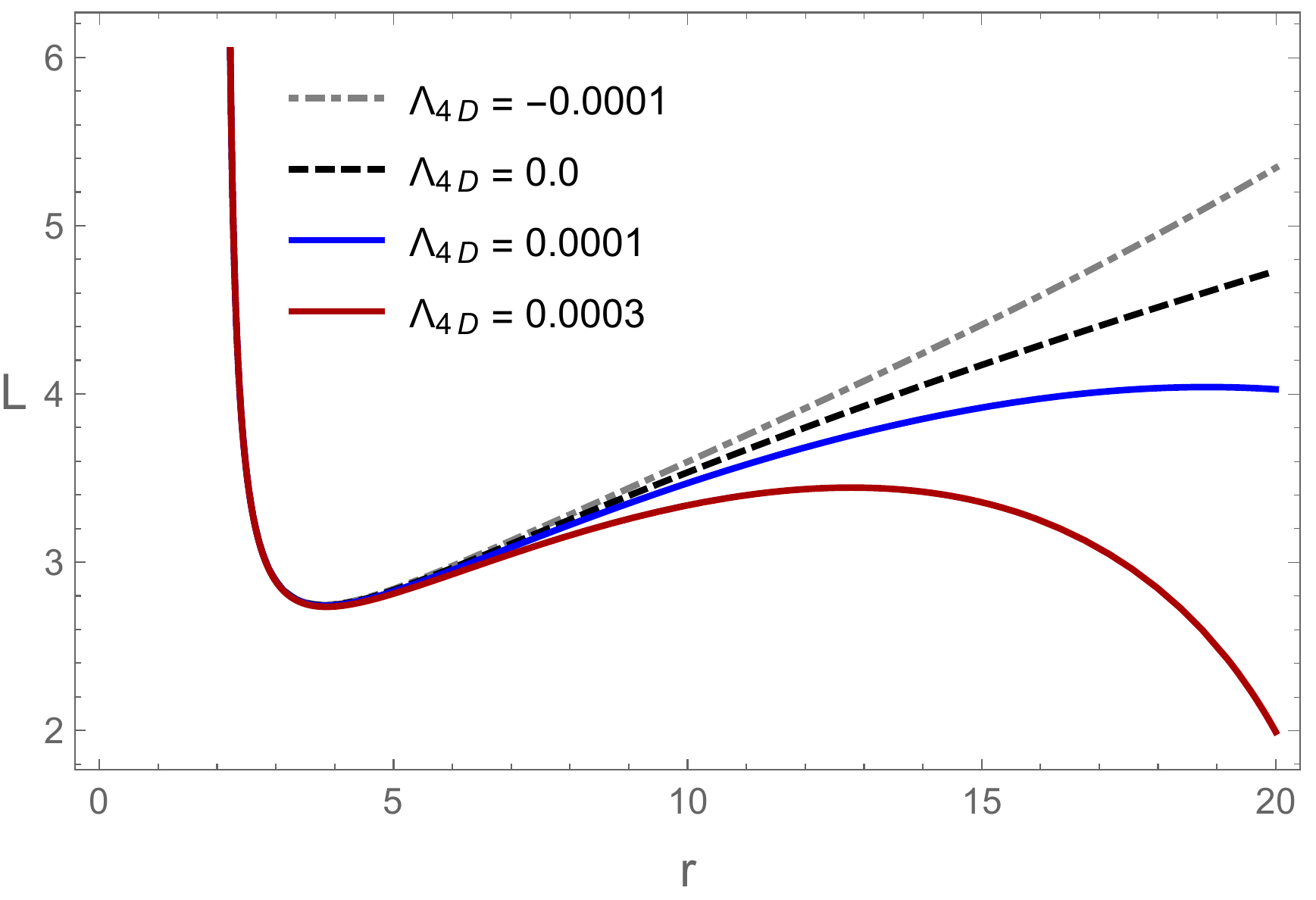}
\end{minipage}
\begin{minipage}[b]{0.58\textwidth}\hspace{-1.1cm}
\includegraphics[width=0.8\textwidth]{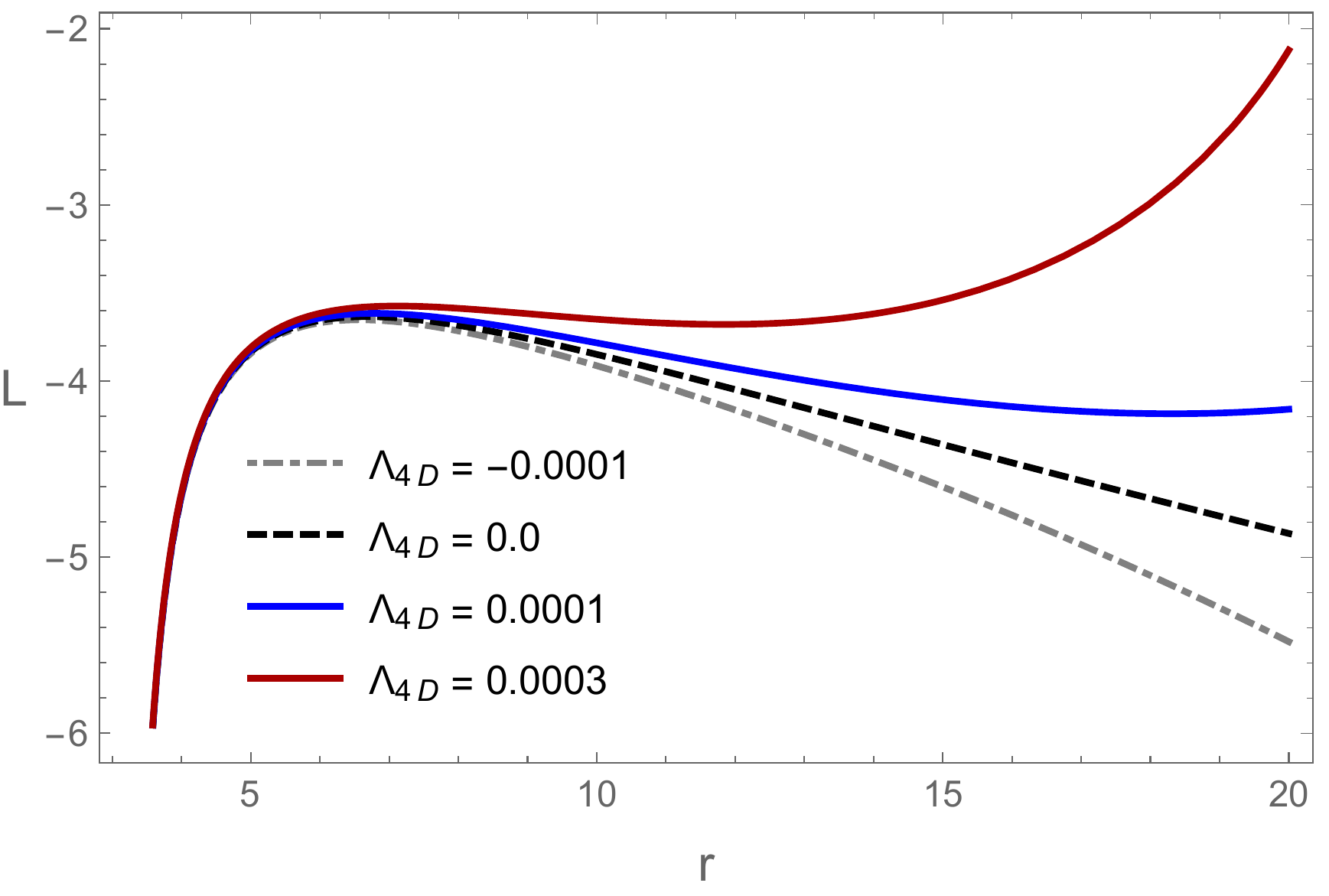}
\end{minipage}
\caption{Radial profile of the circular orbit's angular momentum, left for prograde and right for retrogade particles at $a=0.4$ and $\Lambda_{4D}=0.0001$ (top row), while at $a=q=0.4$ (bottom row).}\label{AM}
\end{figure*}
%%----------------------------------------------------%%
The radii of photon orbits can be obtained by considering the equality of Eq. \eqref{G21}. The radial profile of energy and angular momentum of the circular orbits near a braneworld BH with a cosmological constant is shown in Figs. \ref{Energy} and \ref{AM}. The innermost stable circular orbits can be determined by minima of the radial profile.
Graph describes the affected behaviour of $\mathcal{E}$, under the influence of tidal charge $q$ and a cosmological constant $\Lambda_{4D}$. From where we conclude that $q<0$ contribute to the energy, while $q>0$ diminishing the radial profile of $\mathcal{E}$ for both prograde and retrograde particles. Similar behaviour is observed for the cosmological constant where $\Lambda_{4D}<0$ increase the energy, shown by the grey dashed-dotted curves in Fig. \ref{Energy}, while $\Lambda_{4D}>0$ results in a decrease of $\mathcal{E}$ for both prograde and retrograde orbits.
Besides, we observed that in case of $q<0$, the angular momentum increases for direct orbit, while a decrease in the case of indirect orbits and similar behaviour of the angular momentum is obtained for $\Lambda_{4D}<0$. On the other hand, the negative value of $\Lambda_{4D}$ contributes to the stability of stable orbits ($L>0$), while destabilizes it in the case of counter-rotating particles ($L<0$). Furthermore, the positive value of $\Lambda_{4D}$ diminishing the angular momentum of prograde orbit and increase it in the case of retrograde particles motion at a larger radial distance $r$.  
%%------------------------------------------------------%%
\subsection{Effective potential}
%%------------------------------------------------------%%
The concept of effective potential ($U_{eff}$) is very important in BH physics, as it can be used to determine the range of angular momentum and to discuss the stability of circular orbits. The stable and unstable circular orbits can be acquired with the help of maximum and minimum values of $U_{eff}$, respectively. The radial Eq. \eqref{G6}, for $\Xi L \neq a \mathcal{E}$, can be expressed as
\begin{equation}\nonumber
\frac{1}{2}\dot{r}^{2}+U_{eff}(r)=0.
\end{equation}
%%-----------------------------------------------------%%
\begin{figure*}
\begin{minipage}[b]{0.58\textwidth} \hspace{-0.2cm}
\includegraphics[width=0.8\textwidth]{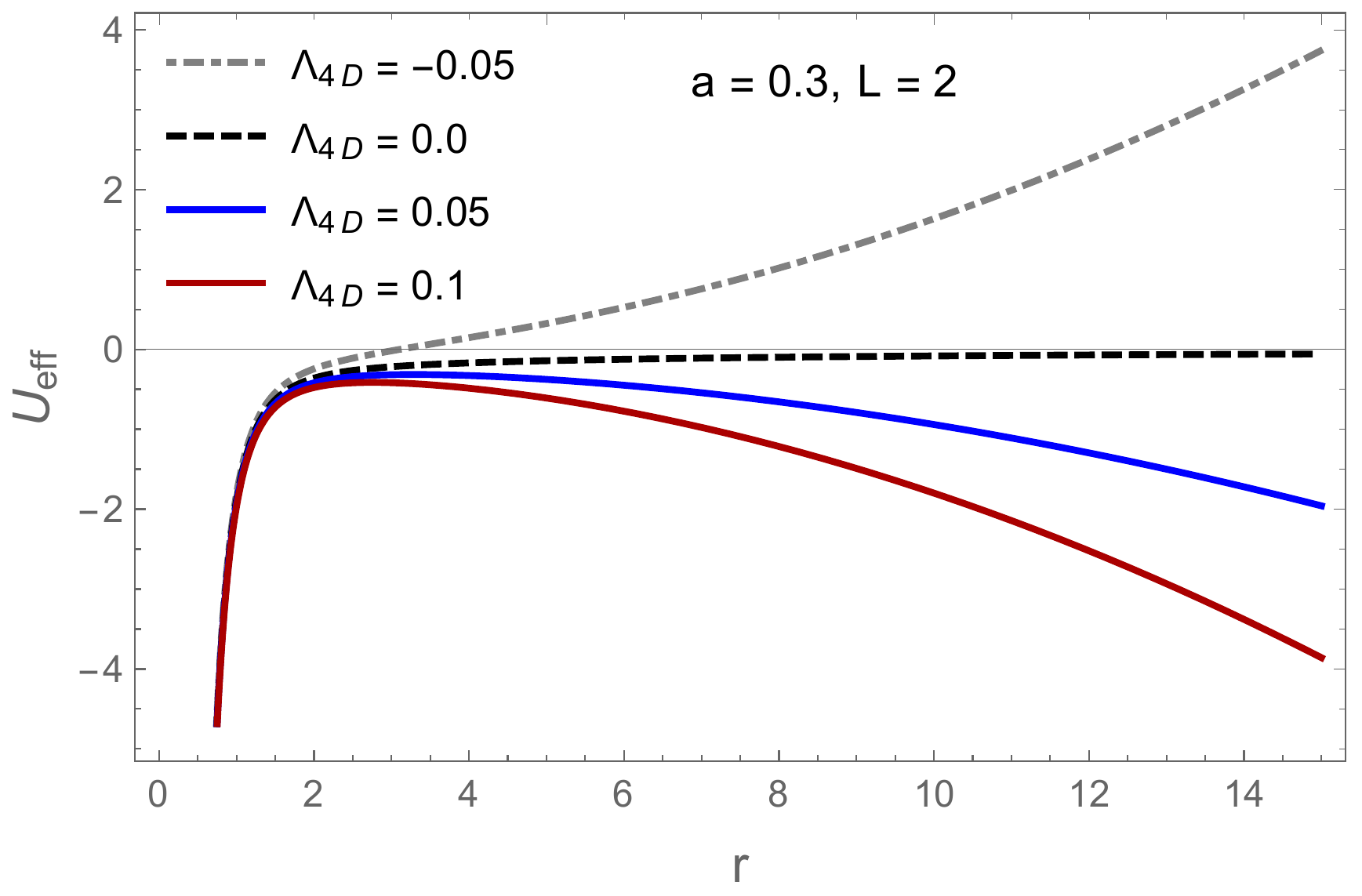}
\end{minipage}\vspace{0.3cm}
\begin{minipage}[b]{0.58\textwidth} \hspace{-1.1cm}
\includegraphics[width=0.8\textwidth]{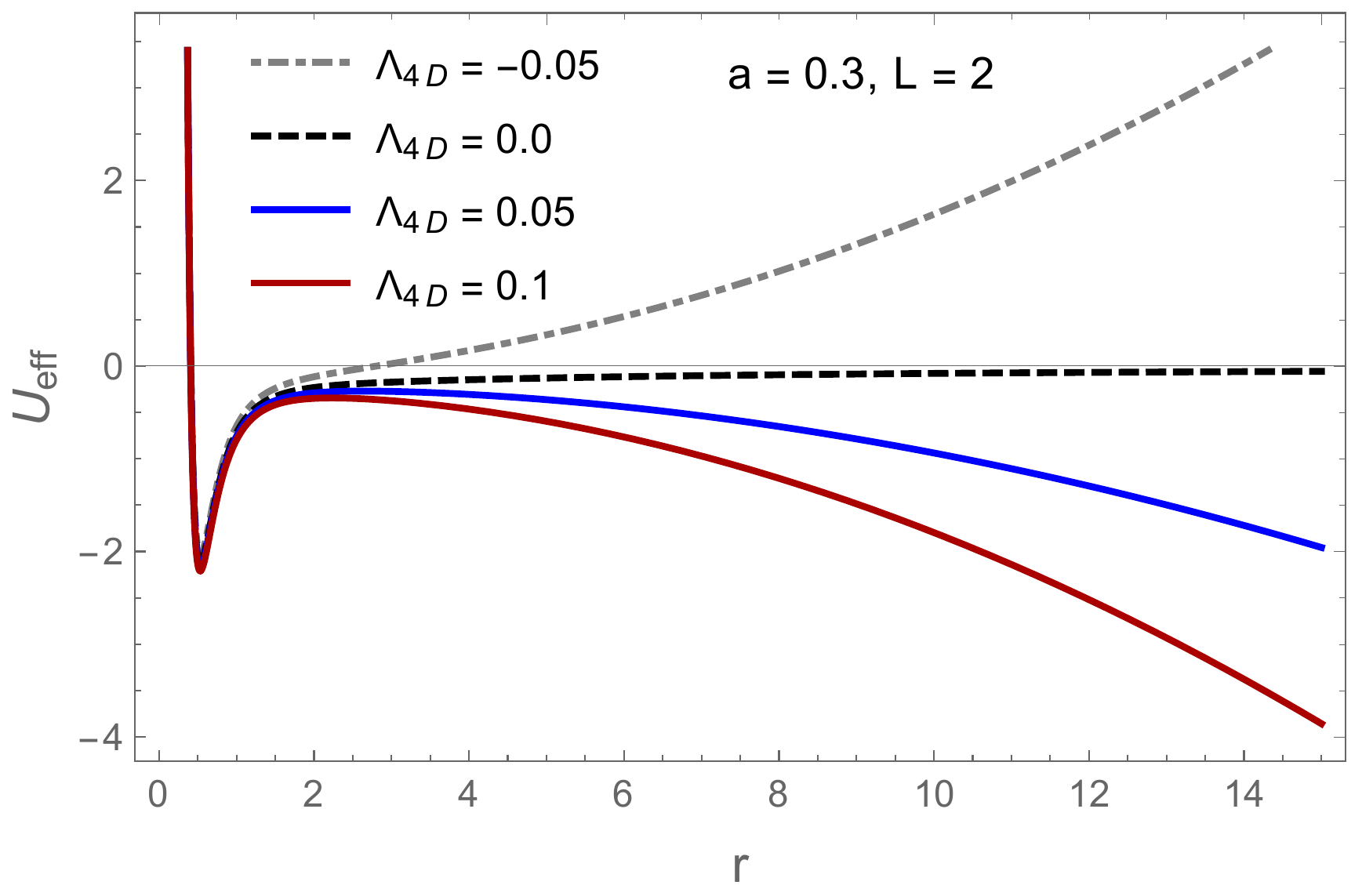}
\end{minipage}
      %%----------%%
\begin{minipage}[b]{0.58\textwidth} \hspace{-0.2cm}
\includegraphics[width=0.8\textwidth]{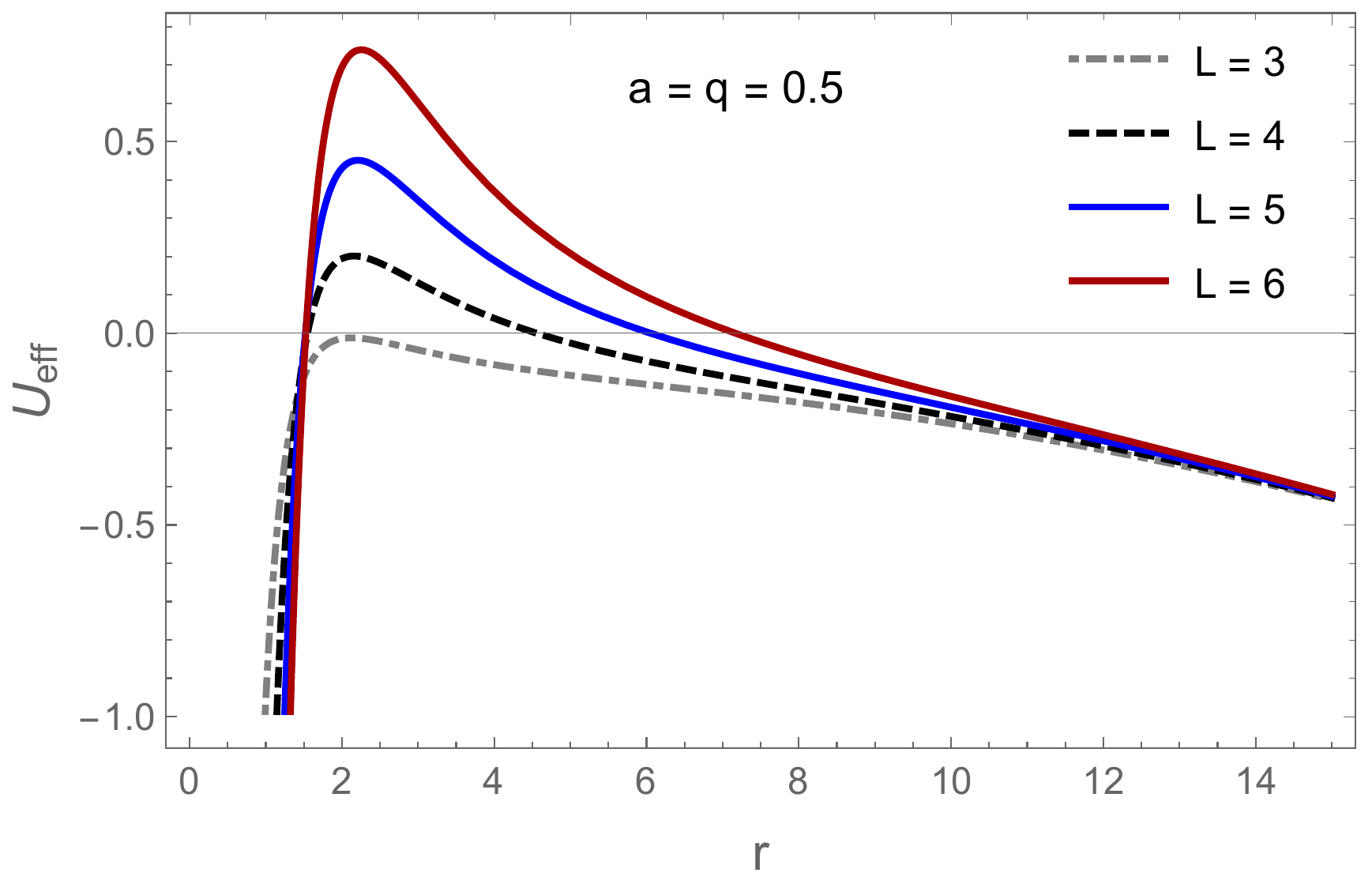}
\end{minipage}
\begin{minipage}[b]{0.58\textwidth}\hspace{-1.1cm}
\includegraphics[width=0.8\textwidth]{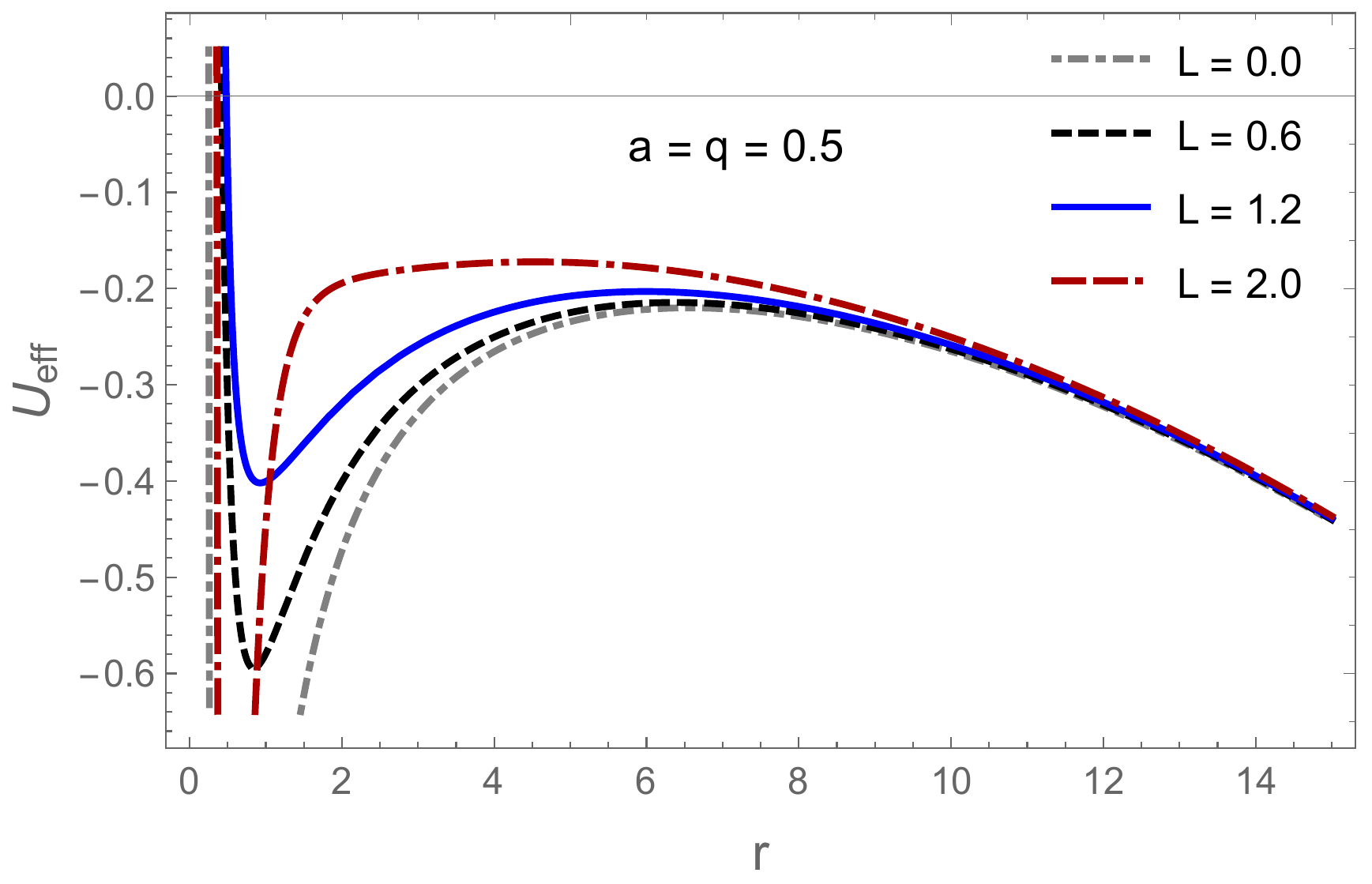}
\end{minipage}
\caption{The plotted behaviour of effective potential vs $r$, top row at $q=0$ (left) and $q=0.5$ (right), whereas the bottom row is plotted for $\Lambda_{4D}=0$ (left) and $\Lambda_{4D}=0.01$ (right).}\label{EP1}
\end{figure*}
%%----------------------------------------------------%%
Hence, for the timelike particle along geodesics, $U_{eff}$ takes the form
\begin{equation} \label{eff1}
U_{eff}=\frac{1}{2r^4}\left(\Delta_r \left(r^2+(\Xi L-a\mathcal{E})^2 \right)- \left((a^2+r^2)\mathcal{E}-\Xi aL\right)^2 \right).
\end{equation}
The initial radial acceleration, as well as the radial velocity, should be vanished for a particle in the circular orbits, i.e.,
 \begin{eqnarray*}
\frac{\partial U_{eff}}{\partial r}=0, \quad U_{eff}=0.
\end{eqnarray*}
In case of the stable circular orbit, the effective potential should be minimum
\begin{equation*}
\frac{\partial{^2} U_{eff}}{\partial r^{2}}>0.
\end{equation*}
%%-----------------------------------------------------%%
\begin{figure*}
\begin{minipage}[b]{0.58\textwidth} \hspace{-0.2cm}
\includegraphics[width=0.8\textwidth]{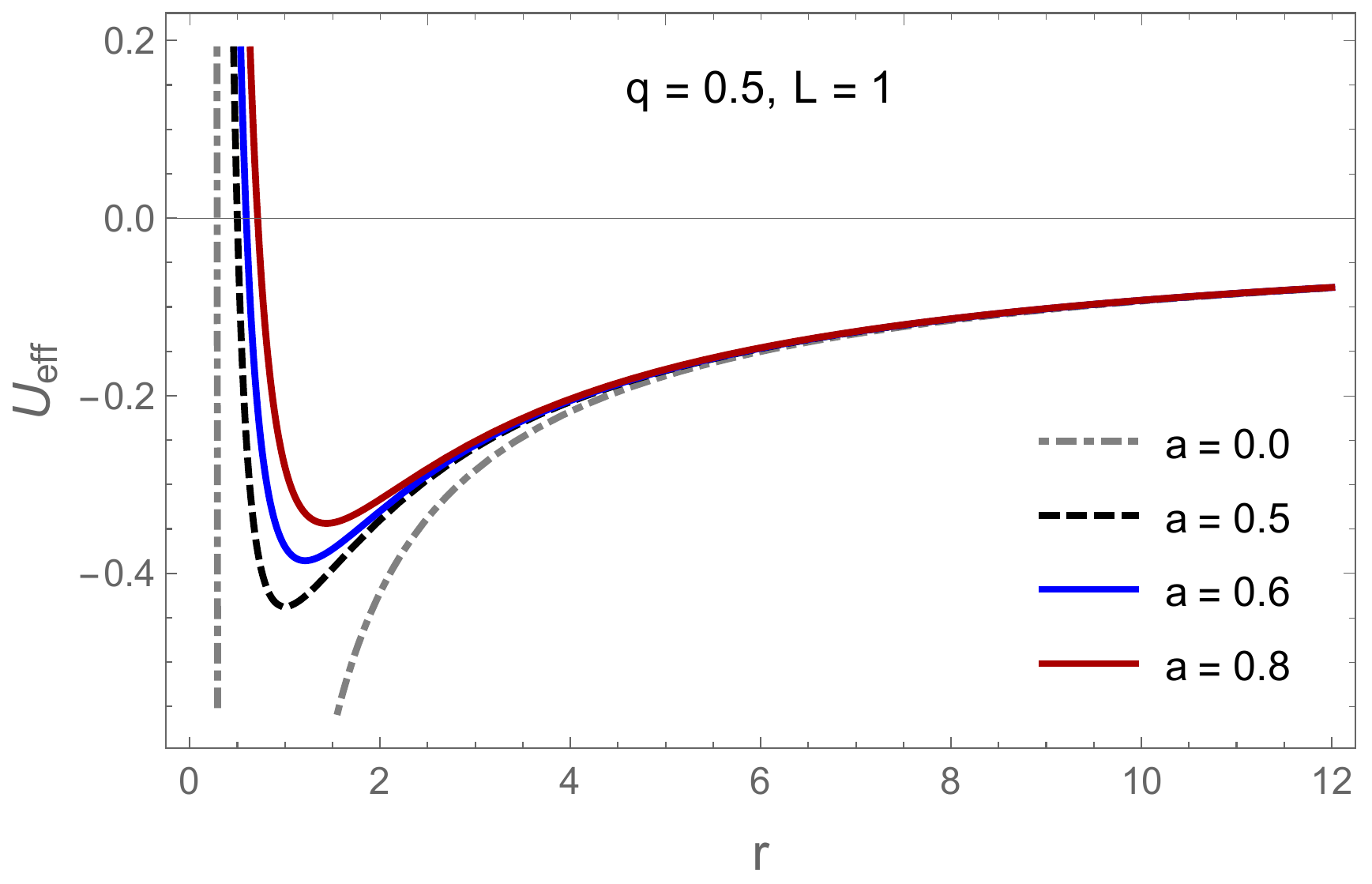}
\end{minipage}\vspace{0.3cm}
\begin{minipage}[b]{0.58\textwidth}\hspace{-1.1cm}
\includegraphics[width=0.8\textwidth]{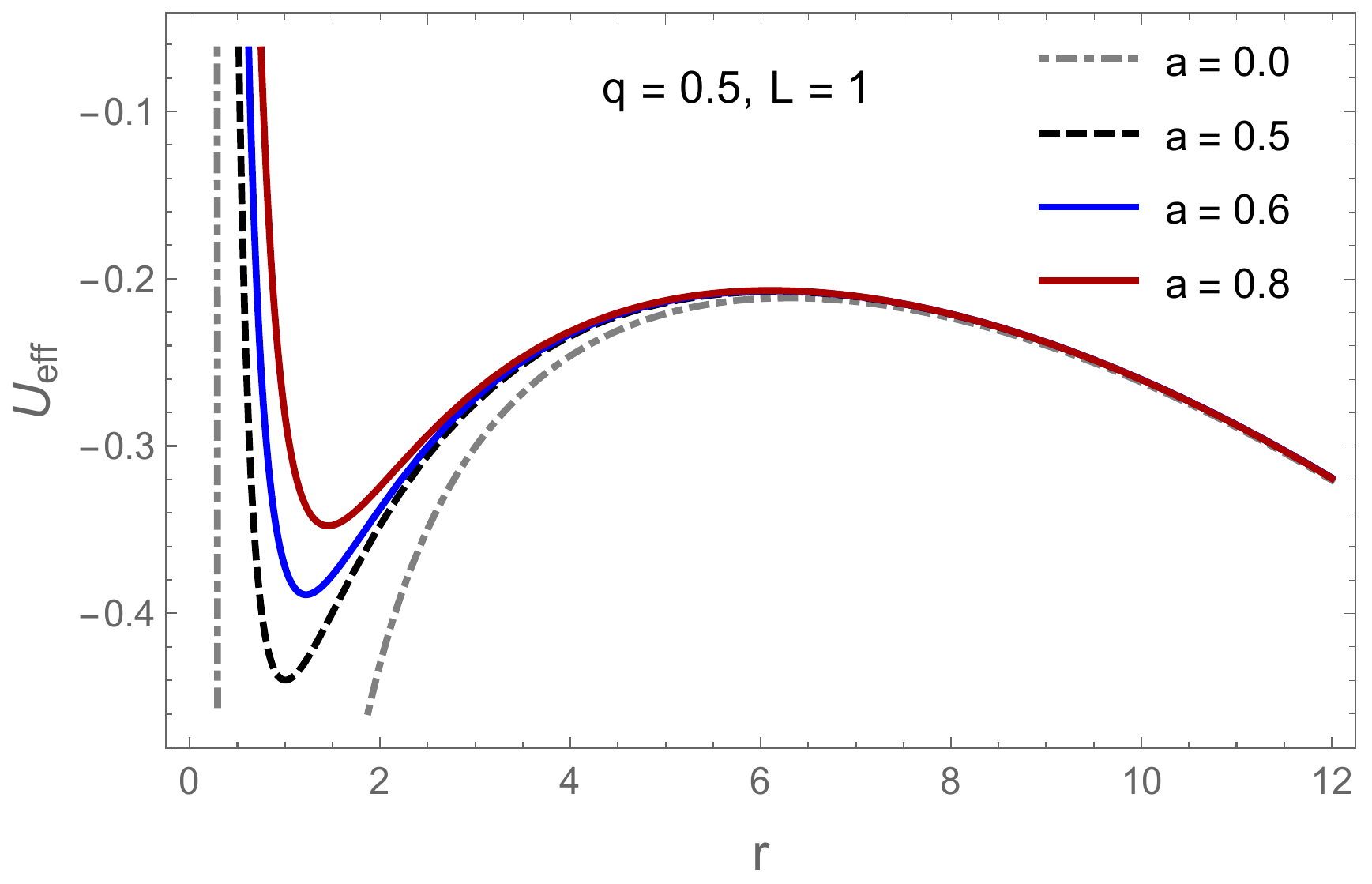}
\end{minipage}
\caption{The plotted behaviour of effective potential vs $r$, at $\Lambda_{4D}=0.0$ (left) and $\Lambda_{4D}=0.01$ (right).}\label{EP2}
\end{figure*}
%%----------------------------------------------------%%
The radial dependence of $U_{eff}$ for the test particle motion around a rotating braneworld BH with a cosmological constant is plotted in Figs. \ref{EP1} and \ref{EP2}. The graphical behaviour shows that the negative value of cosmological constant results in stabilizing the circular orbits. On the other hand, it can clearly be observed that $\Lambda_{4D} > 0$, leads us to the unstable circular orbits. Whereas, the variation of angular momentum shows that grater value of $L$ ends up with unstable circular orbits. The graph shows that $L$ increases the instability of circular orbits and the orbits attain its maximum values  $U_{eff} \approx -0.01,0.20,0.46,0.74$, at $L=3, 4, 5, 6$, respectively near $r=2$. Moreover, at a smaller value of $L$, $U_{eff}$ is stable but becomes unstable at a larger radial distance $r$. BH spin appears to increase the stability of circular orbits but in the presence of a cosmological effect, the stable orbits become unstable at a larger radial distance $r$ Fig. \ref{EP2} (right).
%%-----------------------------------------------------%%
\section{Shadow}
%%-----------------------------------------------------%%
In general, BH's shadow could be described as the boundary of photon capture and scattering orbits \cite{Chandrasekhar}. Since an observer at a special infinity can distinguish the shadow of a BH over a bright background, as a dark region. In the case of stationary BHs, the shadow appeared to be a perfect circular disc, whereas BH rotation cause distortion to the shadow of rotational BHs. On considering null geodesics and by following the Hamiltonian-Jacobi approach, the radial and azimuthal components of the geodesics equation can be obtained as \cite{Eiroa}
\begin{eqnarray}\label{BS5}
\mathcal{R}(r)&=&\left((a^2+r^2)\mathcal{E}-\Xi a L_z\right)^2-\Delta_r \left((a\mathcal{E}-L)^2+\mathcal{O} \right),
\\\label{BS6} \Theta(\theta)&=& -\left(a \mathcal{E} \sin^2\theta-\Xi L \right)^2 \csc^2\theta-\Delta_\theta\left((a\mathcal{E}-L)^2+\mathcal{O}\right).
\end{eqnarray}
Here $\mathcal{O}=\mathcal{K}-(a\mathcal{E}-L)^2$, denotes Carter's constant of separation. The above equations describe photon orbits of a spinning braneworld black hole with a cosmological constant. In principle, they depend on two impact parameters described in terms of constant of motion $\mathcal{E}$ and $L$, as $\xi=L/\mathcal{E} $ and $\eta=\mathcal{O}/\mathcal{E}^2$ \cite{Chandrasekhar}. In the case of photons, there exist three types of geodesics called scattering, spherical and plumbing orbits; and in terms of $\xi$ and $\eta$ the radial Eq. \eqref{BS5}, could be expressed as
\begin{equation}\label{BS12}
\mathcal{R}(r)=\frac{1}{\mathcal{E}^2}\left[ \left((a^2+r^2)-\Xi  a\xi \right)^2 -\Delta_r\left((a-\xi)^2+\eta \right) \right].
\end{equation}
The effective potential of photons can be obtained from Eq. \eqref{BS12}. As a result, the circular unstable and critical orbits could be acquired using the maximum effective potential, obeying the following conditions
\begin{eqnarray}\label{BS13}
\mathcal{R}(r)=0=\frac{\partial \mathcal{R}(r)}{\partial r}.
\end{eqnarray}
Let us suppose that both observer and photons are situated at a special infinity. Henceforth, using the constraints in Eq. \eqref{BS13}, we can calculate the celestial coordinates ($\xi$, $\eta$) as
\begin{eqnarray}\nonumber\label{BS14}
\xi&=&\frac{1}{{a \left(a^2 \Lambda_{4D}+3\right)\left(r \left(a^2 \Lambda_{4D}+2 r^2\Lambda_{4D}-3\right)+3 M\right)}}\\ 
&&\times [9 a^2(M+r)+3a^2 \left(a^2+r^2\right)r\Lambda_{4D}+9 r (r (r-3 M)+2 q)],\\\nonumber\label{BS15}
\eta&=&\frac{1}{a^2 \left(a^2 \Lambda_{4D}+3\right)^2 \left(r \left(a^2\Lambda_{4D}+2r^2 \Lambda_{4D} -3\right)+3M\right)^2} [-81 r^2(4a^2(q-M r)\\\nonumber
&&+(r(r-3 M)+2q)^2)+54a^2 r (2 a^4 (M+r)-r^3(r(3M + r)- 2q) \\\nonumber
&&+a^2\left(-3M^2 r +2M \left(q-2r^2\right)+2qr + 3r^3\right))\Lambda_{4D} +9a^4(a^4 (-M^2+2Mr\\\nonumber&&
+7r^2)+8a^2 r^2 (r(r-2 M)+q)+r^4 (8q-r (12M+5r)))\Lambda_{4D}^2-6a^6 r\\
&&\left(a^4(M-r)+a^2 r^2(2M + r)+ 4r^5\right)\Lambda_{4D}^3 -a^8 r^2 \left(a^2+2r^2\right)^2\Lambda_{4D}^4].
\end{eqnarray} 
On substituting $q=\Lambda_{4D} = 0$, the  Eqs. \eqref{BS14} and \eqref{BS15}, exactly reduces to the case of Kerr BH
\begin{eqnarray}\label{BS16}
&&\xi=\frac{a^2 (r+M)+r^2 (r-3 M)}{a (M-r)},\\
&&\eta=-\frac{r^3 \left(r (r-3 M)^2-4 a^2 M\right)}{a^2 (M-r)^2}.
\end{eqnarray}
The aforementioned impact parameters are essential, as they determine the shadow boundary. The shape of shadow cast by a BH could be ideally visualize using celestial coordinates and can be described as \cite{Hioki,Chandrasekhar}
\begin{eqnarray}\label{BS17}
&&{\alpha}= \lim_{r_{0} \to \infty} \left(-r_0^2 \sin\vartheta \frac{d\phi}{dr} \right),\\\label{BS18}
&&{\beta}= \lim_{r_{0} \to \infty}\left(r_0^2 \frac{d\phi}{dr} \right).
\end{eqnarray}
In Eqs. \eqref{BS17} and \eqref{BS18}, $r_{0}$ shows the distance from an observer to the BH, whereas $\theta$ represents the angle of inclination between observer’s line of sight and BH rotational axis. Using geodesics equations, the above celestial coordinates takes the form
\begin{eqnarray}\label{BS19}
&&{\alpha}= -{\xi}{\csc \vartheta},\\\label{BS20}
&&{\beta}=\pm \sqrt{\eta+a^2\cos^2\vartheta-\xi^2 \cot^2\vartheta}.
\end{eqnarray}
On considering the equatorial plane ($\theta=\pi/2$), Eqs. \eqref{BS19} and \eqref{BS20} simplifies to
\begin{eqnarray}\label{BS21}
&&{\alpha}=-\xi,\\\label{BS22}
&&{\beta}=\pm \sqrt{\eta}.
\end{eqnarray}
%%-----------------------------------------------------%%
\begin{figure*}
\begin{minipage}[b]{0.58\textwidth} \hspace{-0.2cm}
        \includegraphics[width=0.8\textwidth]{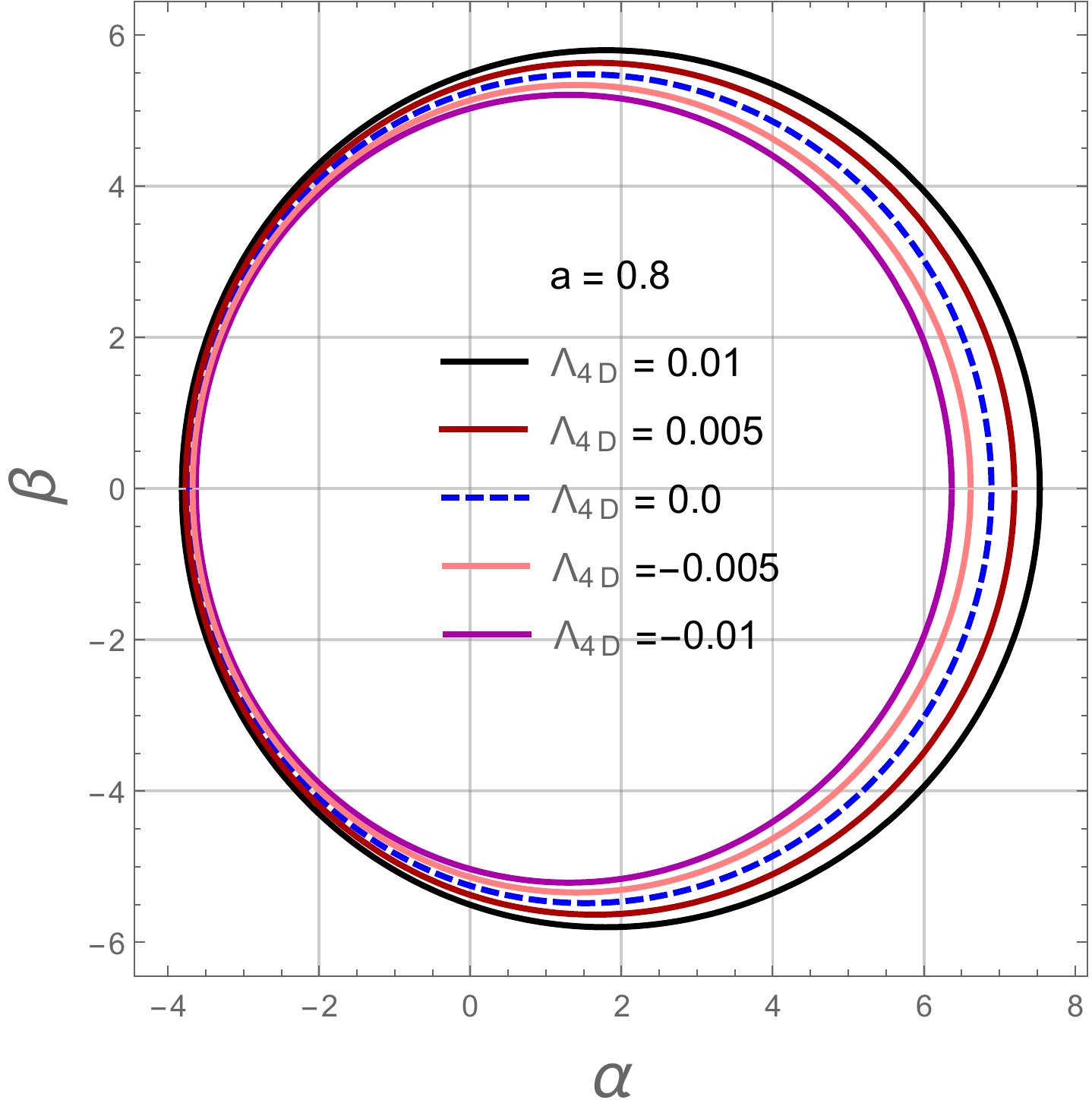}
    \end{minipage}%\vspace{0.3cm}
        \begin{minipage}[b]{0.58\textwidth} \hspace{-1.0cm}
       \includegraphics[width=.8\textwidth]{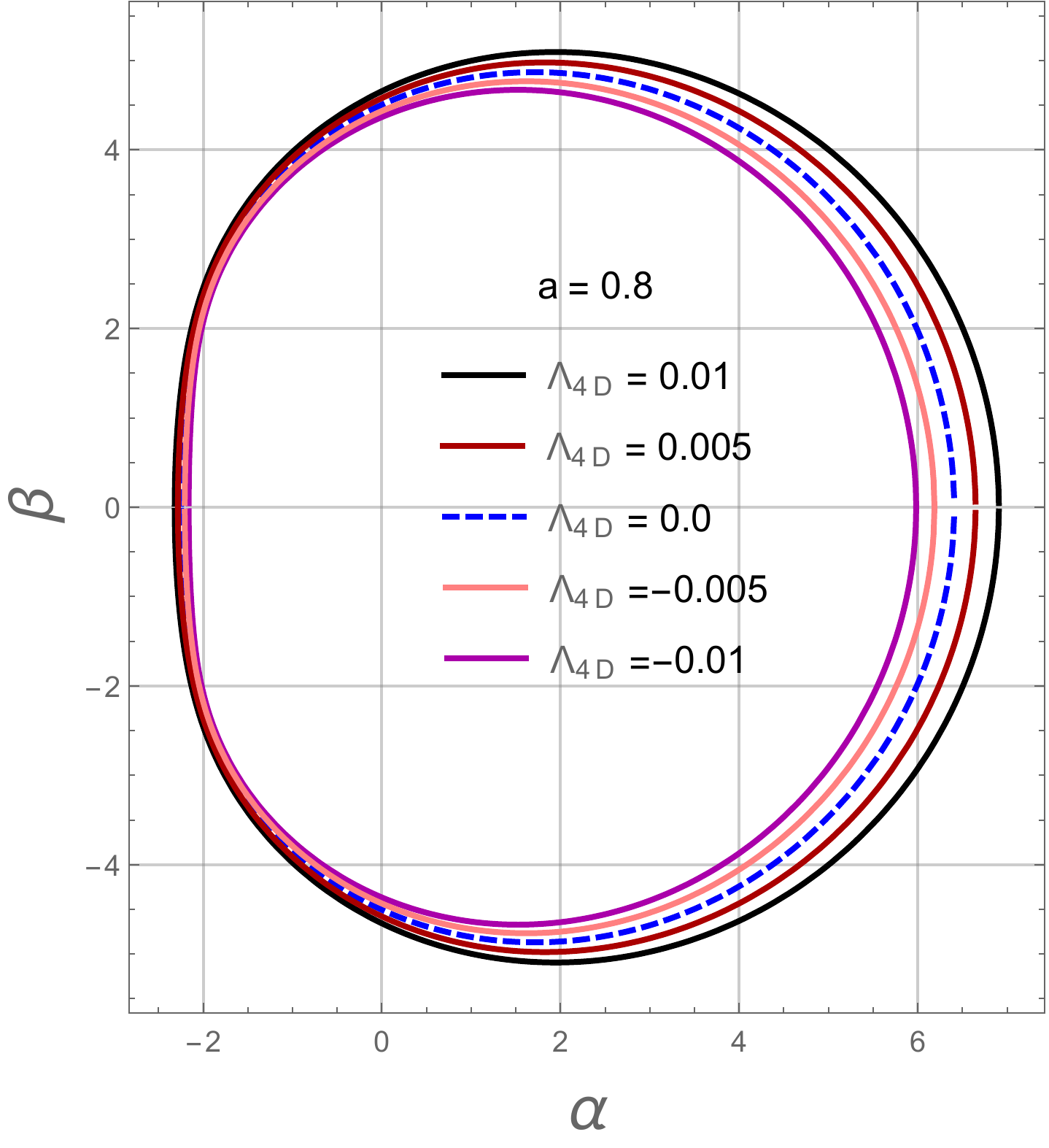}
    \end{minipage}
\caption{Shadow cast by a rotating braneworld BH with a cosmological constant at $q=-0.35$ left, while at $q=0.35$ right panel.}\label{Shadow1}
\end{figure*}
%%-----------------------------------------------------%%
\begin{figure*}
\begin{minipage}[b]{0.58\textwidth} \hspace{-0.2cm}
        \includegraphics[width=0.8\textwidth]{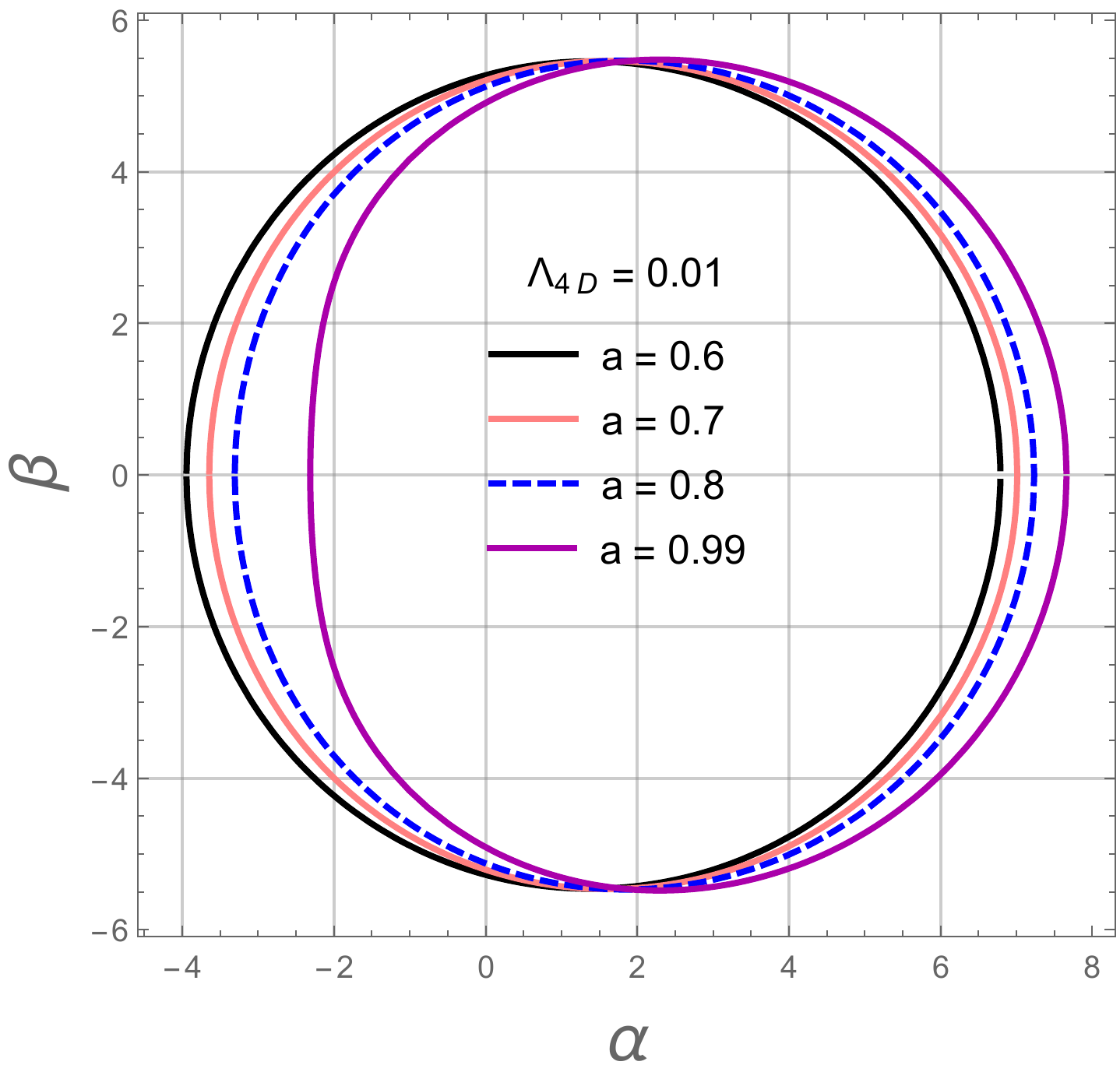}
    \end{minipage}\vspace{0.2cm}
        \begin{minipage}[b]{0.58\textwidth} \hspace{-1.0cm}
       \includegraphics[width=.8\textwidth]{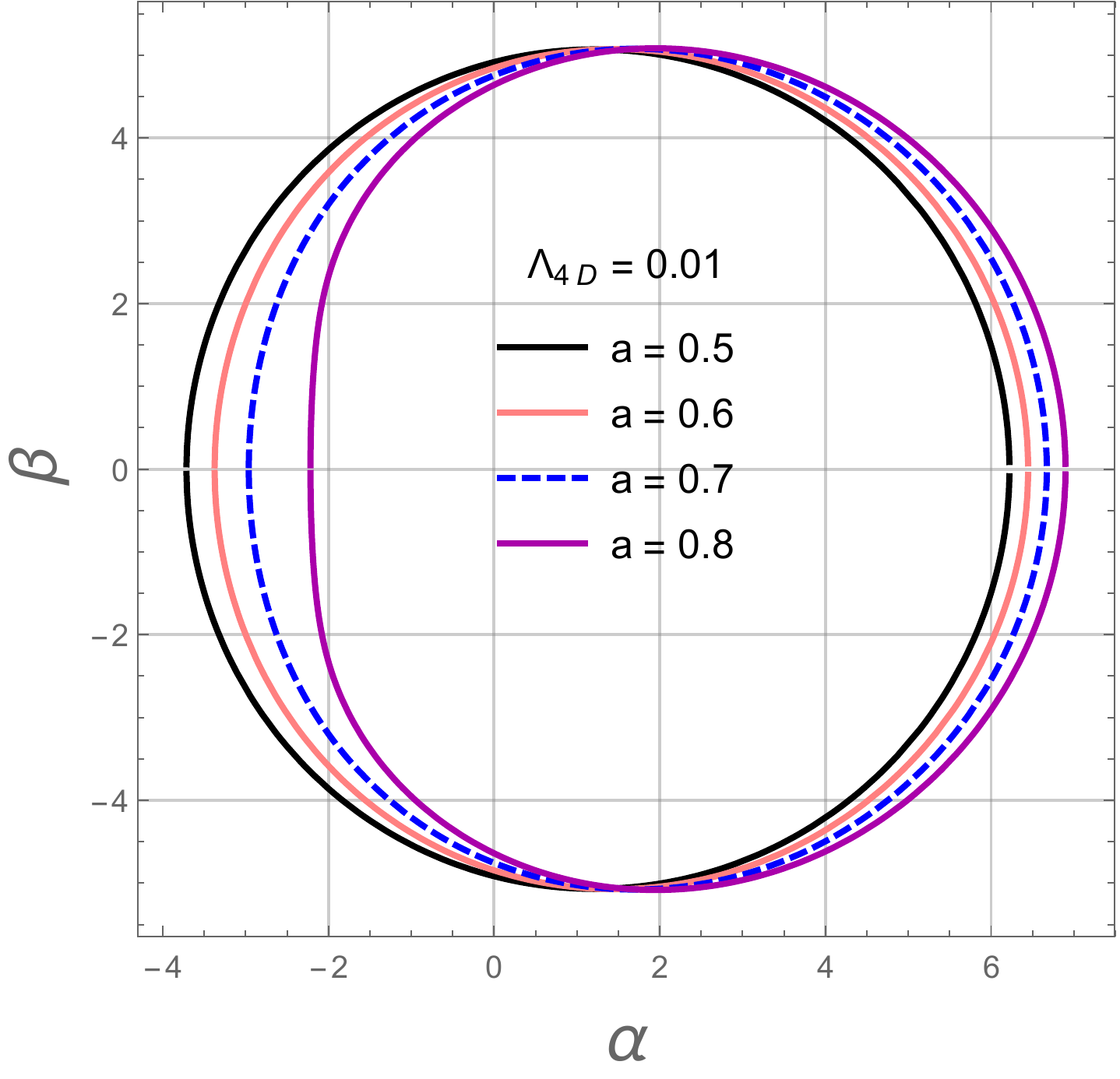}
    \end{minipage}
    %%-------------------%%
    \begin{minipage}[b]{0.58\textwidth} \hspace{-0.2cm}
        \includegraphics[width=0.8\textwidth]{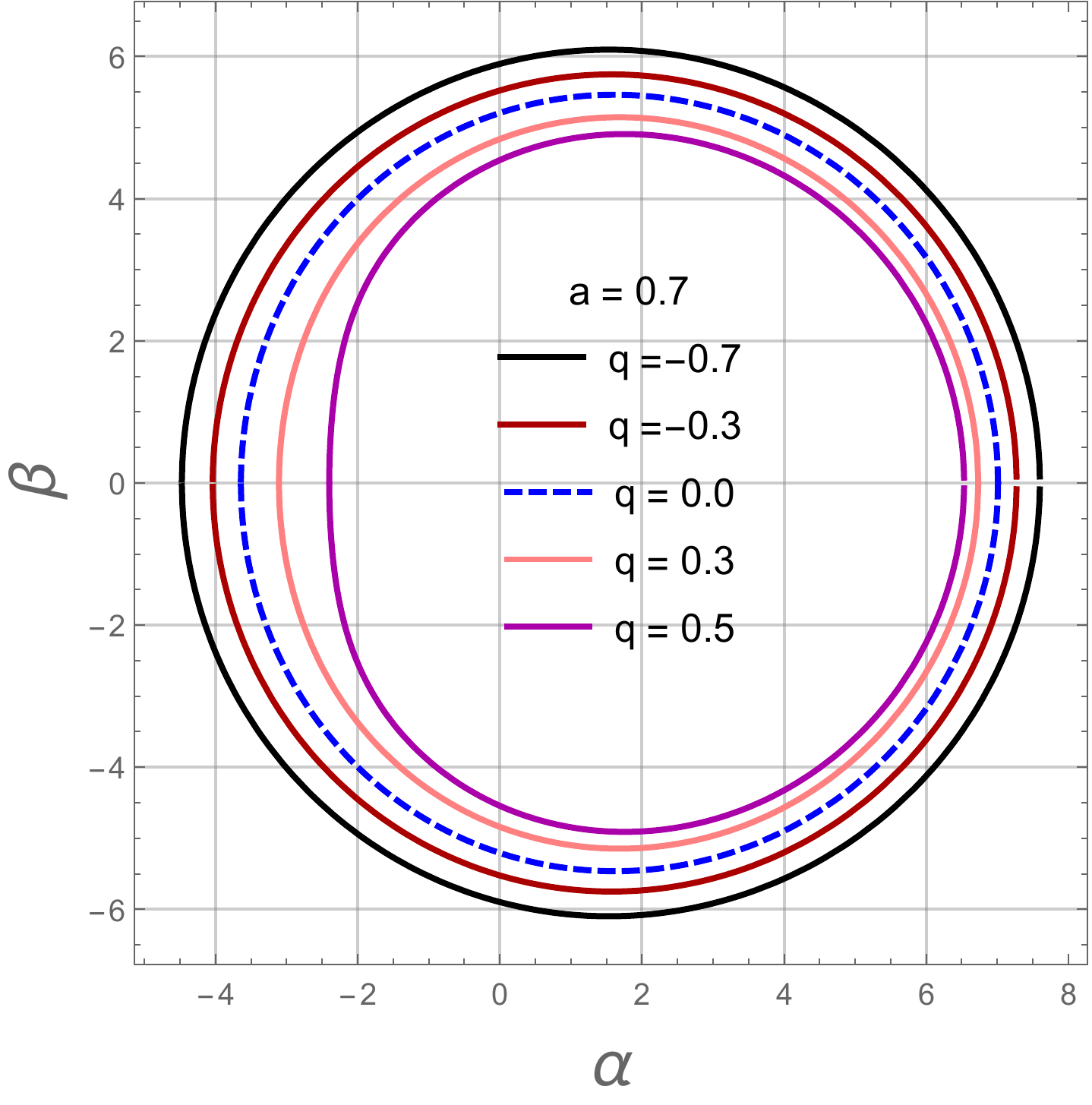}
    \end{minipage}
        \begin{minipage}[b]{0.58\textwidth} \hspace{-1.0cm}
       \includegraphics[width=.8\textwidth]{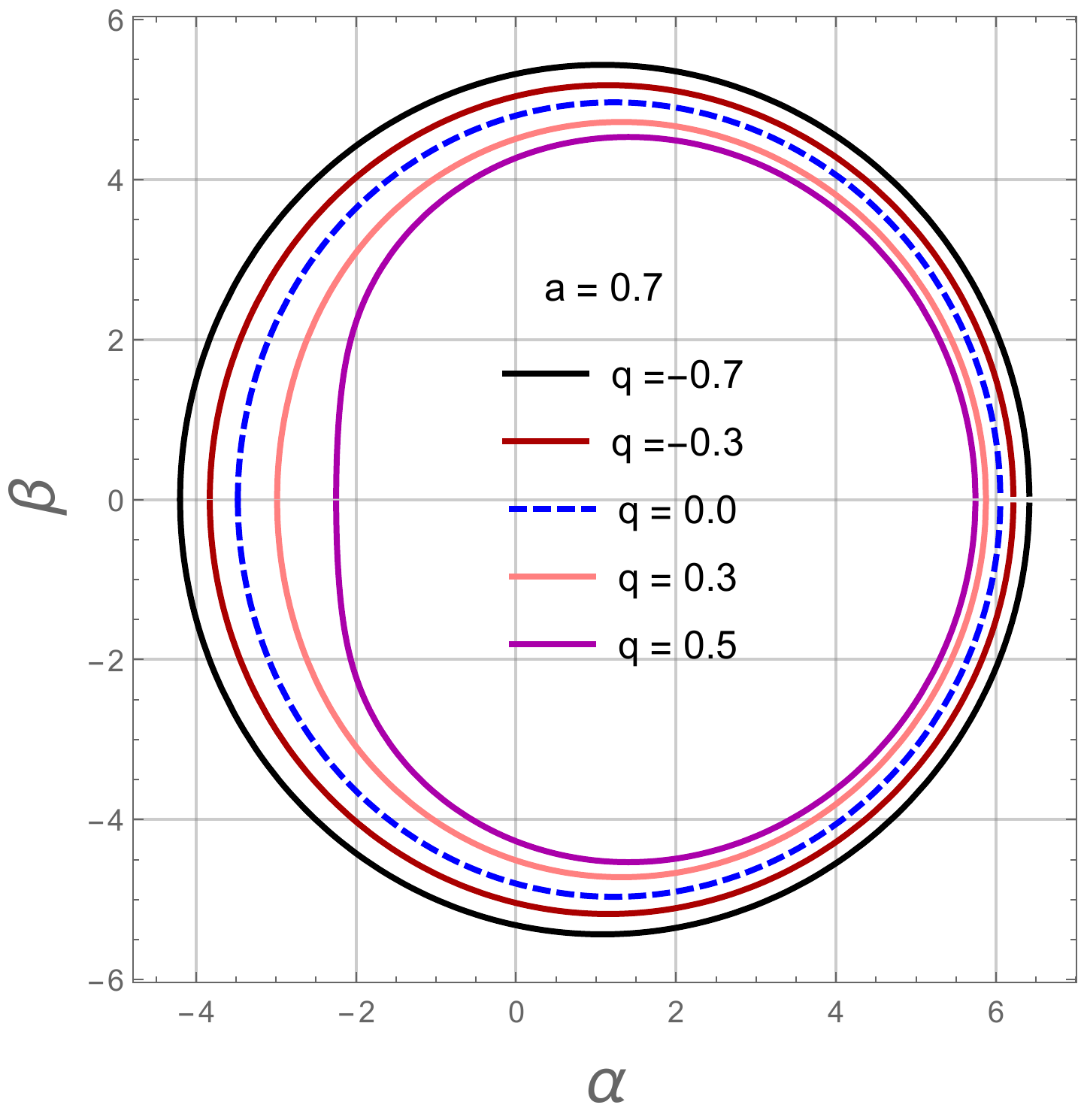}
    \end{minipage}
    \caption{Shadow of the said BH, in the top row at $q=0$ (left) and $q=0.36$ (right), whereas in the bottom row left channel is plotted for $\Lambda_{4D}=-0.01$ and the right one is for $\Lambda_{4D}=0.01$.}\label{Shadow2}
    \end{figure*}
%%-----------------------------------------------------%%
\begin{figure*}
\begin{minipage}[b]{0.58\textwidth} \hspace{-0.2cm}
        \includegraphics[width=0.8\textwidth]{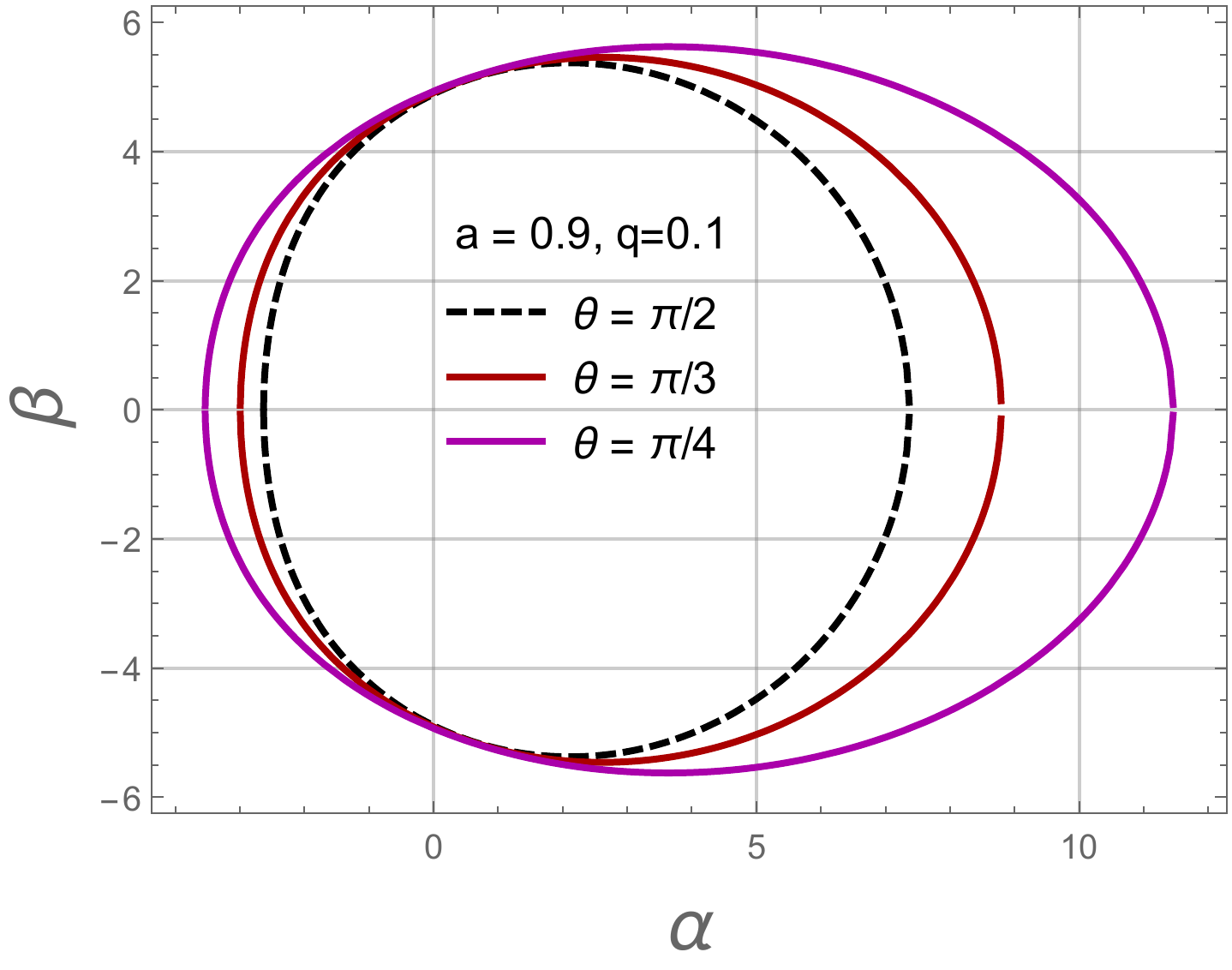}
    \end{minipage}\vspace{0.cm}
        \begin{minipage}[b]{0.58\textwidth} \hspace{-1.0cm}
       \includegraphics[width=.8\textwidth]{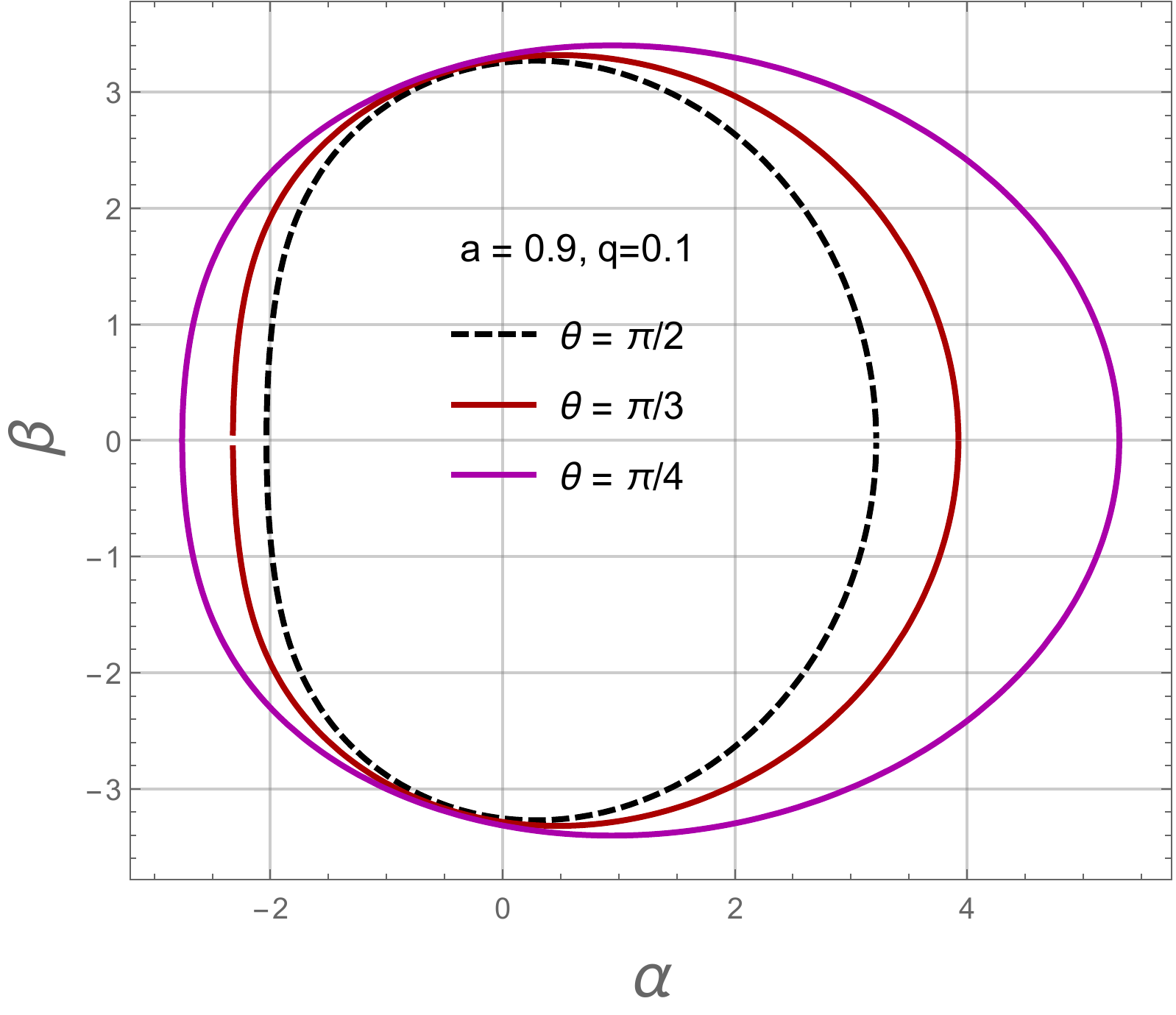}
    \end{minipage}
    \caption{Shadow cast for different values of the inclination angles $\theta$, at $\Lambda_{4D}=0.01$ (left) and $\Lambda_{4D}=-0.015$ (right).}\label{Shadow3}
    \end{figure*}
%%%-----------------------------------------------------%%
The numerical investigation of shadow cast by a braneworld BH with a cosmological constant under the influence of spacetime parameters is plotted in Figs. \ref{Shadow1}, \ref{Shadow2} and \ref{Shadow3}. The graphical illustration shows that negative value of the cosmological constant increase the radius of BH shadow, whereas its positive value results in a decrease of the shadow radius. Similar behaviour of the apparent shadow is observed for the tidal charge $q$ of the BH. Moreover, our finding shows that in response to the dragging effect, BH's rotation elongates its image towards the rotational axis, which coincides with the earlier findings \cite{Ovgun1,Medeiros}. While both BH rotation and positive tidal charge distorts BH shadow and its distortion becomes maximum in case of faster rotating BHs. Fig. \ref{Shadow3} demonstrate the apparent shape of BH shadow at different values of the inclination angle $\theta$.
%%%----------------------------------------------------%%
\subsection{Observables}
\label{sec:51}
%%-----------------------------------------------------%%
In order to illustrate the apparent shape of a rotating braneworld BH with a cosmological constant, we introduce two observables, $R_s$ (shadow radius) and $\delta_s$ (distortion parameter) \cite{Hioki}. The shadow radius $R_s$ is approximated by a reference circle's radius that passes through  three points on shadow's boundary, while on the other hand $\delta_s$ outlines the deviation of shadow from the circle.
Henceforth, $R_s$ and $\delta_s$ can be described by the following expressions
\beq
R_s=\frac{(\alpha_t-\alpha_r)^2+\beta_t^2}{2 \mid \alpha_t-\alpha_r \mid}, \quad \delta_s=\frac{\mid \tilde{\alpha_l}-\alpha_l \mid}{R_s}.
\eeq
Here ($\alpha_t, \beta_t$) and ($\alpha_r, \beta_r$), respectively represent the topmost and rightmost points where the reference circle cut the shadow. On the other hand ($\alpha_l,0$) and ($\tilde{\alpha_l},0$), respectively denotes the locations, at which the shadow and reference circle meets the leftmost $\alpha$ axes. 
%%%------------------Energy Emission-------------------%%
\subsection{Energy Emission}
%%%-----------------------------------------------------%%
According to our assumption, for a distant observer at infinity BH's shadow reaches the high energy absorption cross-section of the BH \cite{Misner, Wei}. In case of a spherical symmetric BH, the limiting constant value of the absorption cross-section can be defined as
\begin{equation}\nonumber
\sigma_{ilm} \approx \pi R_s^2.
\end{equation}
Thus the energy emission rate can be computed by making use of the  limiting constant value as,
\begin{equation}\label{ee1}
\frac{d^2E(w)}{dwdt}= \frac{2\pi^2 R_s^2}{\exp^{w/T-1}}.
\end{equation}
Here $w$ denote photon's frequency, while $T$ is the Hawking temperature of the BH at the event horizon, which can be defined as
\begin{eqnarray}\label{ee2}
T(r_+)& = &\lim_{\theta\to 0, r\to r_+} \frac{\partial r \sqrt{g_{tt}}}{2\pi \sqrt{g_{rr}}} \\ \label{ee3}
&=&-\frac{3 a^2 M +3 r_+ (q-M r_+) + r_+ \left(a^2+r_+^2\right)^2\Lambda_{4D}}{6 \pi  \left(a^2+r_+^2\right)^2}.
\end{eqnarray}
In the above equation, $r_{+}$ denotes the event horizon.
%%-----------------------------------------------------%%
\begin{figure*}
\begin{minipage}[b]{0.58\textwidth} \hspace{-0.2cm}
        \includegraphics[width=0.8\textwidth]{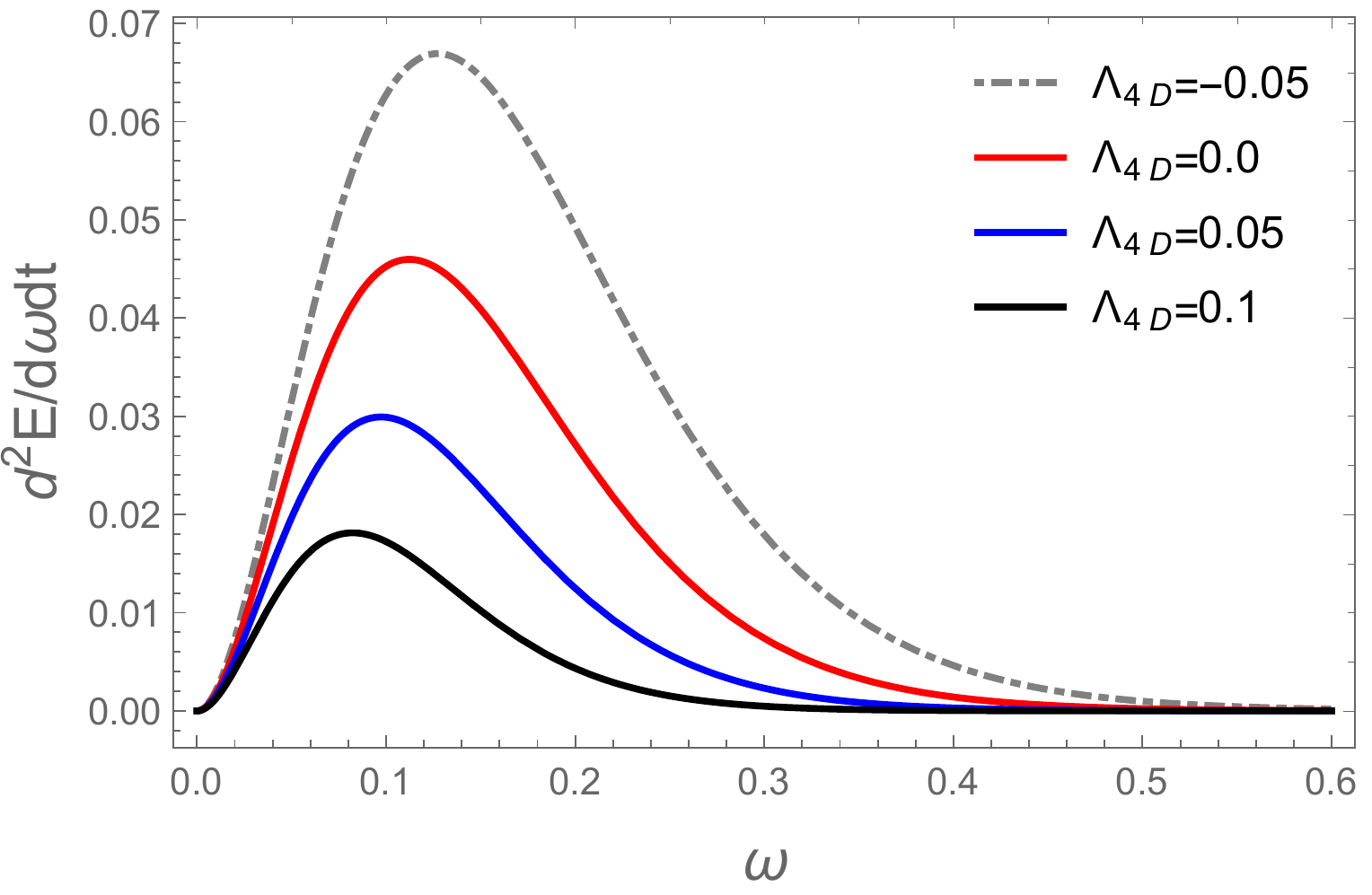}
    \end{minipage}
    \vspace{0.2cm}
        \begin{minipage}[b]{0.58\textwidth} \hspace{-1.2cm}
       \includegraphics[width=.8\textwidth]{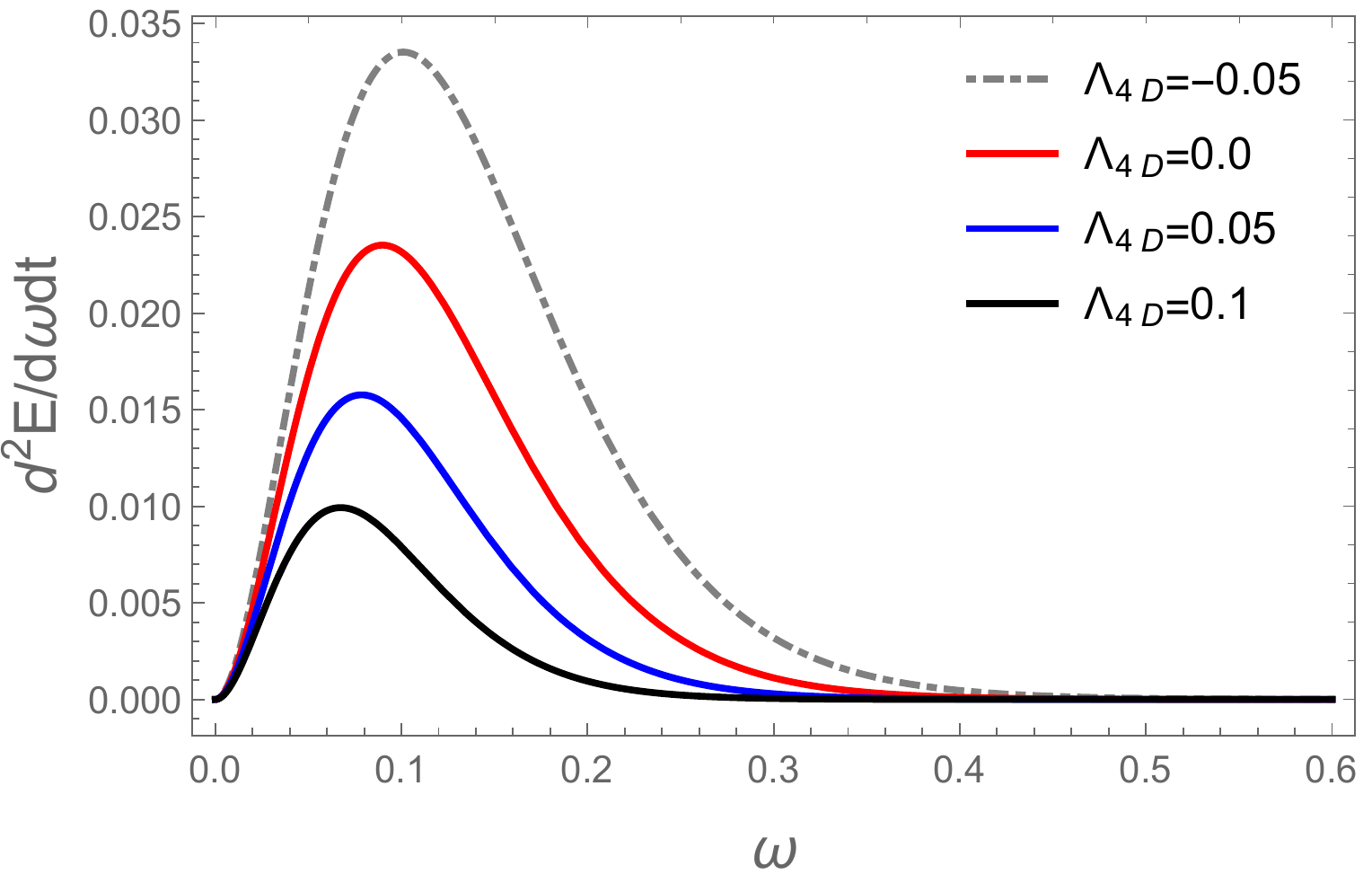}
    \end{minipage}
\begin{minipage}[b]{0.58\textwidth} \hspace{-0.2cm}
        \includegraphics[width=0.8\textwidth]{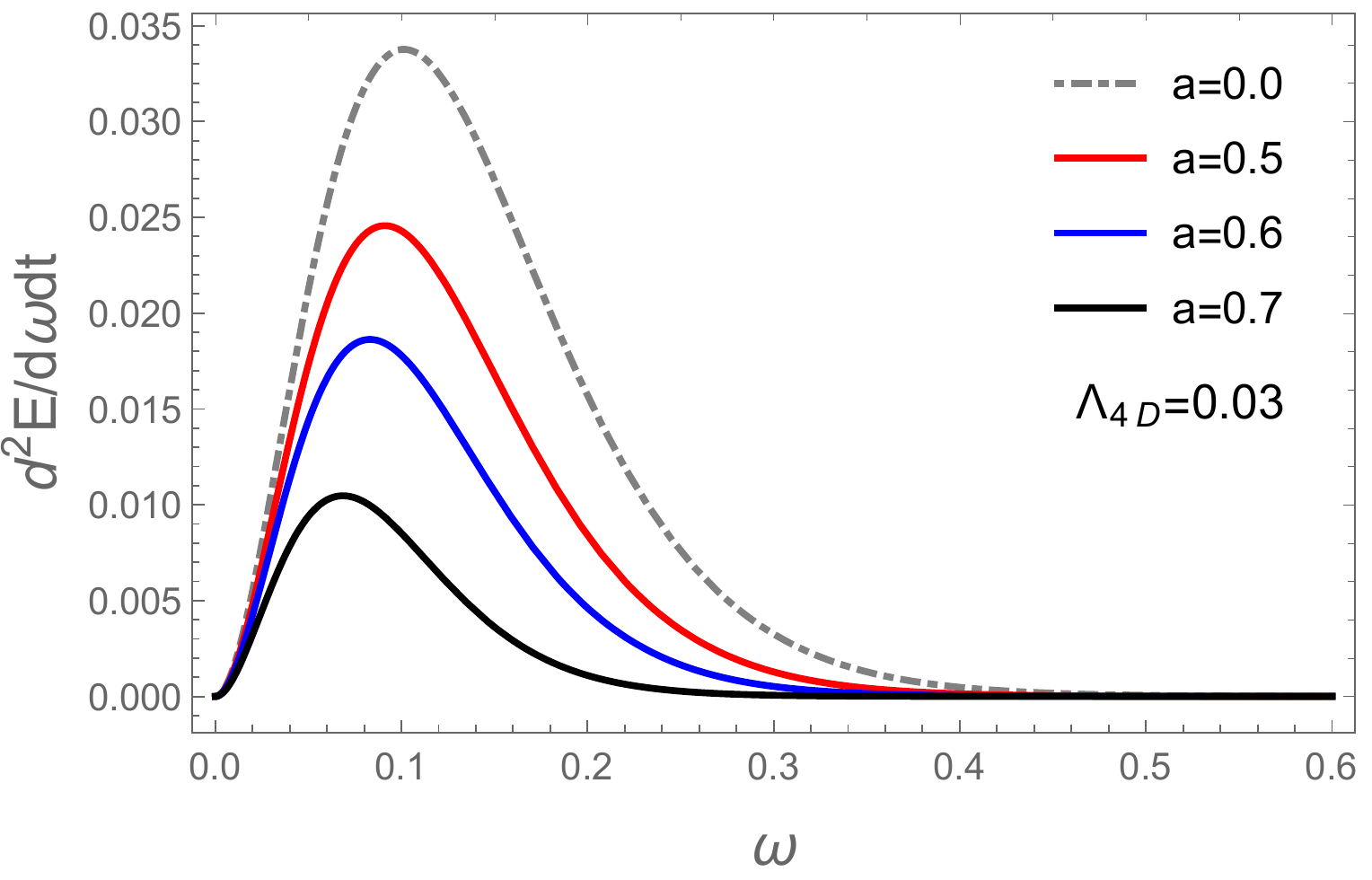}
    \end{minipage}
        \begin{minipage}[b]{0.58\textwidth} \hspace{-1.2cm}
       \includegraphics[width=.8\textwidth]{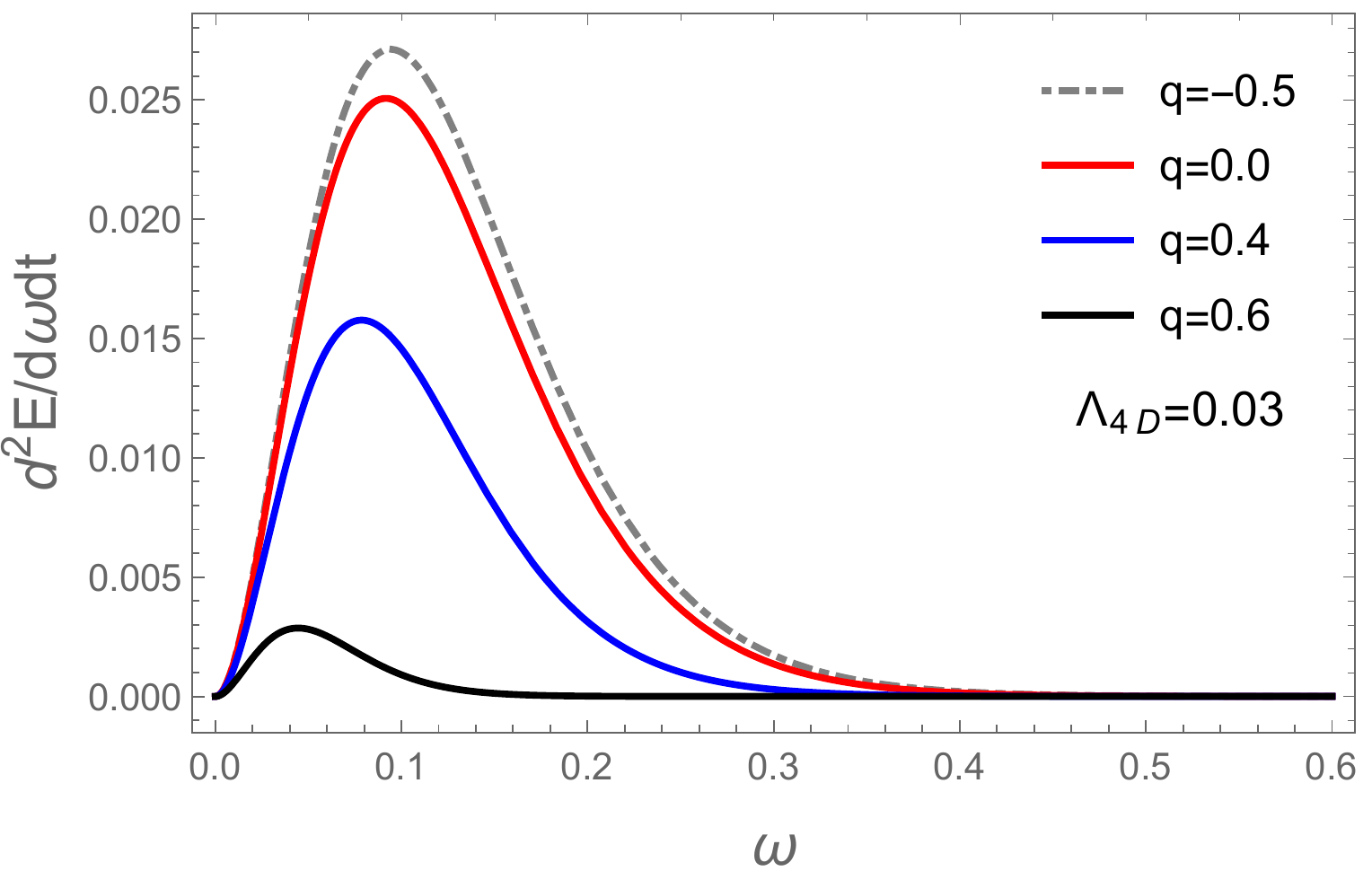}
    \end{minipage}
\caption{The plotted behaviour of energy emission in the upper row at $a=q=0$ left panel and $a=q=0.5$ right panel. The lower row left panel is plotted for $q=0.5$, while the right one is for $a=0.6$.}\label{Eem1}
\end{figure*}
%%-----------------------------------------------------%%
\begin{figure*}
\begin{minipage}[b]{0.58\textwidth} \hspace{-0.2cm}
        \includegraphics[width=0.8\textwidth]{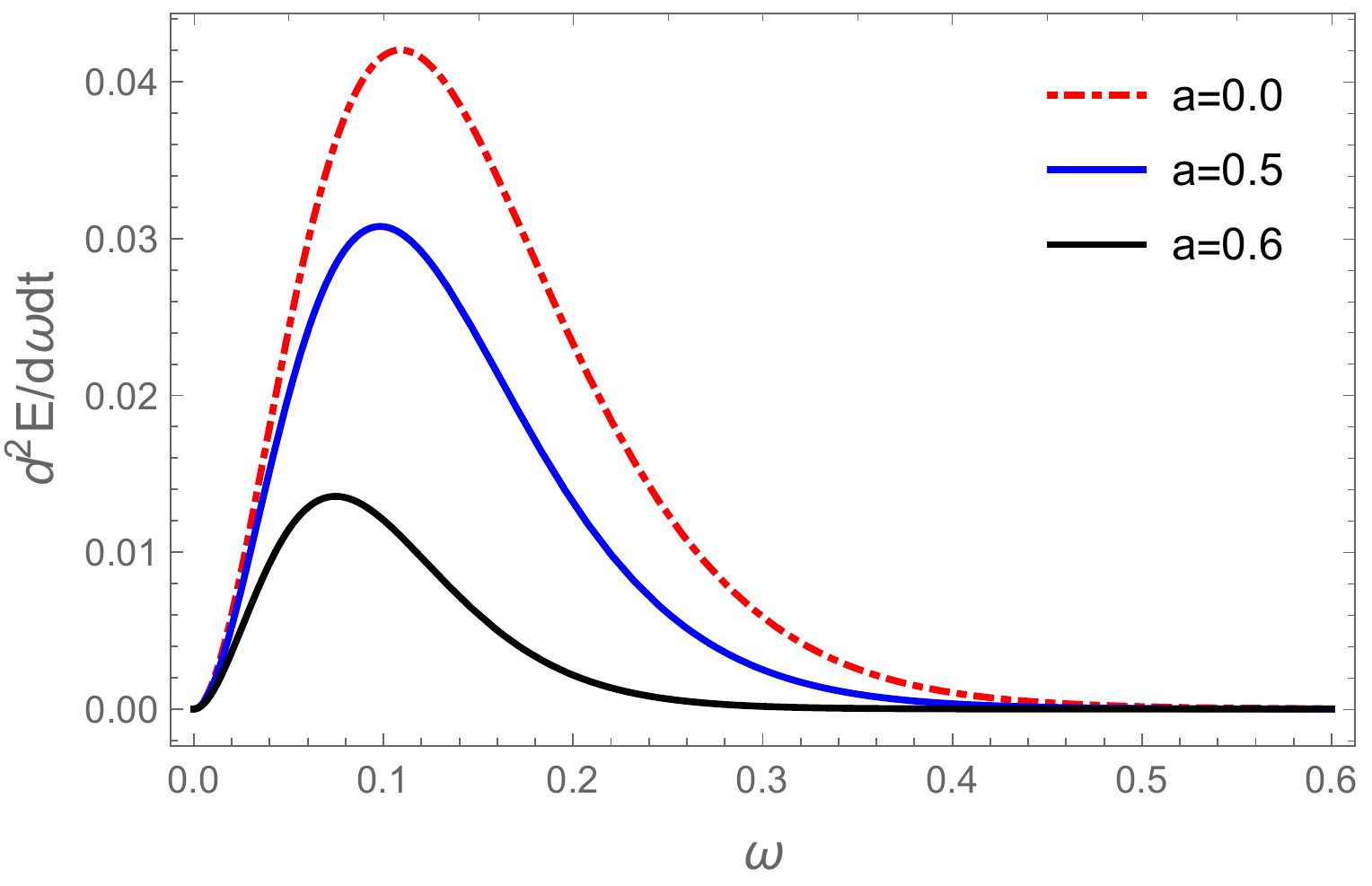}
    \end{minipage}
    \vspace{0.0cm}
        \begin{minipage}[b]{0.58\textwidth} \hspace{-1.2cm}
       \includegraphics[width=.8\textwidth]{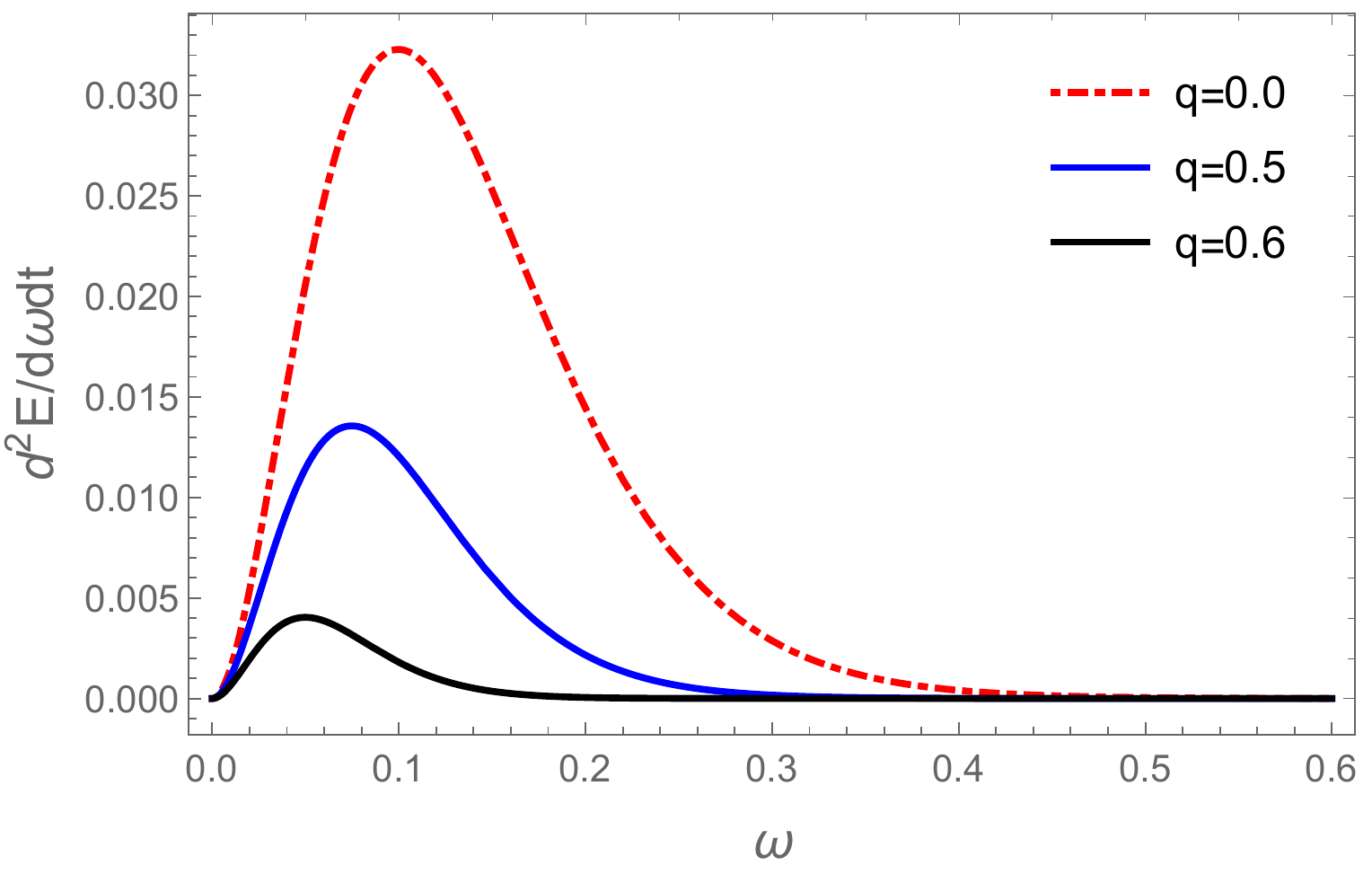}
    \end{minipage}
\caption{The plotted behavior of energy emission rate for $\Lambda=0$, at $q=0.5$ left panel and at $a=0.6$ right panel. 
}\label{Eem2}
\end{figure*}
%%-----------------------------------------------------%%
The Figs. \ref{Eem1} and \ref{Eem2}, describe graphical illustration of the energy emission rate of rotating braneworld BH both with and without a cosmological constant. The graphical behaviour reflects that positive value of the cosmological constant diminishing the rate of energy emission, whereas its negative value results in an increase of the energy emission. On the other hand, both BH spin and brane parameter considerable decreases the rate of energy emission in the presence, as well as in the absence of a cosmological constant. Moreover, we observed that both static and chargeless BHs have more energy emission rate in comparison with spinning and charged BHs.
%%%----------------------------------------------------%%
\section{Conclusion}
\label{sec:5}
%%-----------------------------------------------------%%
In this article, we have studied the circular geodesics and shadow cast by a rotating BH in Randall-Sundrum brane with a cosmological constant. In fact, our primary goal was to establish a theoretical exploration for the possible optical properties of the said BH, that can be used for the future observational investigation, especially for the upcoming EHT observations. We revisited the spacetime geometry of a $4$-dimensional braneworld BH with a cosmological constant and explored the structure of its horizons under the influence of spacetime parameters. It is observed that as far as the BH  rotates faster the area of its event horizons become smaller. The negative value of tidal charge increases the event horizons as well as SLS, while its positive value results in a decrease of the SLS and event horizon. On the other hand, $\Lambda_{4D}>0$ diminishing the event horizon whereas $\Lambda_{4D}>0$ contributes to the area of the event horizon. Besides, the positive value of $\Lambda_{4D}0$ increases the SLS. Furthermore, in comparison with the standard Kerr BH, the braneworld BH with a cosmological constant have a larger area of the event horizon.

To have information on the circular motion, we have considered the timelike geodesics and calculated the expressions for energy, angular momentum and effective potential of particles. From the numerical investigation, we have observed that $\Lambda_{4D}<0$ increases the energy, while $\Lambda_{4D}>0$ diminishing the energy of both co-rotating and counter-rotating particles. In addition, the circular orbits have more energy in the absence of a cosmological constant as compared to the circular orbits with a positive cosmological constant. Our result in the absence of cosmological constant shows considerable resemblance with the acquired results of the regular rotating BH \cite{Toshmatov}. We have observed that $\Lambda_{4D}<0$ results in stabilizing the circular orbits, while $\Lambda_{4D} > 0$, leads us to the unstable circular orbits. Our findings describe that $L$ increases the instability of circular orbits and attains its maximum values  $U_{eff} \approx -0.01,0.20,0.46,0.74$, at $L=3, 4, 5, 6$, respectively near $r=2$. For the variation of $L$, the shape of $U_{eff}$ is analogous to the results of rotating Ayón-Beato-García BHs \cite{Ahmed}. Moreover, for the smaller value of $L$, $U_{eff}$ is stable but becomes unstable at a larger distance $r$. BH spin appears to increase the stability of circular orbits but in the presence of a cosmological effect, the stable orbits become unstable at a larger radial distance $r$.

By making use of the Hamiltonian-Jacobi approach, we have obtained the mathematical expressions of celestial coordinates for the shadow cast by a rotating BH in Randall-Sundrum brane with a cosmological constant. On substituting $q=\Lambda_{4D}=0$, the celestial coordinates reduce to the case of Kerr BH, which validates our analytical expressions \cite{Wei}. Besides, the BH spin our aim was to explore the effects of tidal charge and cosmological constant on BH's shadow. From the numerical investigation, we have observed that in response to the dragging effect, BH's rotation elongates its image toward the rotational axis. While the negative value of tidal charge and cosmological constant contributes to the shadow area, which is consistent  with the exploration of Oliveira \cite{Oliveira}. On the other hand, their positive values result in diminishing the shadow radius analogously to the findings of Banerjee et. al \cite{Banerjee}. Interestingly, effects of the cosmological constant are indistinguishable to the effect of quintessence parameter on BH's shadow presented in our earlier work \cite{Khan3}. We have also shown that at different inclination angles observer observes disparate shapes of the BH shadow. 

Furthermore, our numerical investigation reflects that the positive value of $\Lambda_{4D}$ behaves like the deviation parameter \cite{Amir} and results in diminishing the rate of energy emission, while its negative value contributes to the energy emission. Besides, both BH spin and tidal charge considerably decrease the rate of energy emission. In addition, we noted that both static, as well as chargeless BHs, have grater energy emission rate in comparison with spinning and charged BHs (see Figs. \ref{Eem1} and \ref{Eem2}).
\subsubsection*{Acknowledgment}
This work is supported by the NSFC Project (11771407) and the MOST Innovation Method Project (2019IM050400).
%\begin{acknowledgments}
%J. L. Ren is very grateful for the financial support from the National Nature Science Foundation of China (11771407).
%\end{acknowledgments}
%\nocite{*}
%\bibliography{aipsamp}% Produces the bibliography via BibTeX.

\end{document}